%% file: fhecore.tex
\documentclass[conference]{IEEEtran}

\input{packages}
\newcommand{\fhecore}{FHECore\xspace}
\newcommand{\blue}[1]{\textcolor{blue}{#1}}

\newcommand{\red}[1]{\textcolor{red}{#1}}

\newcommand{\bluehyperlink}[2]{\textcolor{black}{\href{mailto:#2}{#1}}}

\def\BibTeX{{\rm B\kern-.05em{\sc i\kern-.025em b}\kern-.08em
    T\kern-.1667em\lower.7ex\hbox{E}\kern-.125emX}}

\newif\ifarxiv
\arxivtrue          

\begin{document}


\ifarxiv
\title{\fhecore: Rethinking GPU Microarchitecture for Fully Homomorphic Encryption\vspace{-5pt}}
\author{
\IEEEauthorblockA{
Lohit Daksha\IEEEauthorrefmark{1},
Seyda Guzelhan\IEEEauthorrefmark{1},
Kaustubh Shivdikar\IEEEauthorrefmark{5},
Carlos Agulló Domingo\IEEEauthorrefmark{3},
Óscar Vera Lopez\IEEEauthorrefmark{3}, 
Gilbert Jonatan\IEEEauthorrefmark{4},\\
Hubert Dymarkowski\IEEEauthorrefmark{8},
Aymane El Jerari\IEEEauthorrefmark{2}, 
José Cano\IEEEauthorrefmark{8},
José L. Abellán\IEEEauthorrefmark{3},
John Kim\IEEEauthorrefmark{4}, 
David Kaeli\IEEEauthorrefmark{2},
Ajay Joshi\IEEEauthorrefmark{1}
}

\vspace{0.2em}
\IEEEauthorblockA{
\IEEEauthorrefmark{1}Boston University, 
\IEEEauthorrefmark{2}Northeastern University,
\IEEEauthorrefmark{4}KAIST, 
\IEEEauthorrefmark{3}University of Murcia, 
\IEEEauthorrefmark{8}University of Glasgow, 
\IEEEauthorrefmark{5}AMD
}
\vspace{0.2em}
\IEEEauthorblockA{
\{\bluehyperlink{tihol}{tihol@bu.edu},
\bluehyperlink{seyda}{seyda@bu.edu},
\bluehyperlink{joshi}{joshi@bu.edu}\}@bu.edu,
\{\bluehyperlink{eljerari.a}{eljerari.a@ece.neu.edu},
\bluehyperlink{kaeli}{kaeli@ece.neu.edu}\}@ece.neu.edu,
\{\bluehyperlink{gilbertjonatan}{gilbertjonatan@kaist.ac.kr},
\bluehyperlink{jjk12}{jjk12@kaist.ac.kr}\}@kaist.ac.kr,\\
\{\bluehyperlink{carlos.a.d}{carlos.a.d@um.es},
\bluehyperlink{oscar.vera}{oscar.vera@um.es},
\bluehyperlink{jlabellan}{jlabellan@um.es}\}@um.es,
\{\bluehyperlink{2468988d}{2468988d@glasgow.ac.uk},
\bluehyperlink{Jose.CanoReyes}{Jose.CanoReyes@glasgow.ac.uk}\}@glasgow.ac.uk,
\bluehyperlink{kshivdik@amd.com}{kshivdik@amd.com}
}

\vspace{-25pt}
}

\pagenumbering{gobble}
\thispagestyle{empty}
\pagestyle{empty}

\else
\title{\fhecore: Rethinking GPU Microarchitecture \\ for Fully Homomorphic Encryption}

\author{\normalsize{ISCA 2026 Submission
    \textbf{\#\iscasubmissionnumber} -- Confidential Draft -- Do NOT Distribute!!}}

\pagenumbering{arabic}
\thispagestyle{plain}
\pagestyle{plain}
\fi

\microtypesetup{tracking=false}
\maketitle
\microtypesetup{tracking=true}


\input{src/0-abstract}
\input{src/1-introduction}
\input{src/2-background}
\input{src/3-motivation}
\input{src/4-fhecore}
\input{src/5-ckks-on-fhecore}
\input{src/6-evaluation}
\input{src/7-discussion}
\input{src/8-related-work}
\input{src/9-conclusion}

\input{src/acknowledgements}


\microtypesetup{tracking=false}
\bibliographystyle{IEEEtranS}
\bibliography{refs}

\end{document}

%% file: packages.tex
\usepackage{cite}
\usepackage{amsmath,amssymb,amsfonts}

\usepackage[subtle]{savetrees}

\usepackage{algorithm}
\usepackage{algpseudocode}

\algrenewcommand\algorithmiccomment[1]{\hfill\(\triangleright\)~#1}

\usepackage{enumitem}

\usepackage{graphicx}
\usepackage{tabularx}
\usepackage{xcolor}
\usepackage[table]{xcolor}
\usepackage{textcomp}
\usepackage{xspace}
\usepackage{soul}              
\usepackage[normalem]{ulem}    
\usepackage{booktabs,makecell}
\usepackage{multicol,multirow}
\usepackage{adjustbox}
\usepackage{pifont}
\usepackage{tikz}

\usepackage[hyphens]{url}
\usepackage{comment}
\usepackage{fancyhdr}          

\usepackage{hyperref}
\usepackage{placeins}
\usepackage{lipsum}            

\newcommand{\rot}[1]{\rotatebox[origin=c]{90}{#1}}
\newcommand{\cmark}{\ding{51}} 
\newcommand{\xmark}{\ding{55}}

\definecolor{highlightcol}{RGB}{220, 235, 255}  

\definecolor{lohitpurple}{HTML}{CCCCFF}

\definecolor{lilac}{RGB}{147,146,232}

\definecolor{darkred}{RGB}{191,4,4}

\newcommand{\green}[1]{\textcolor{green!60!black}{#1}}

\sethlcolor{yellow!40}

\newcommand{\CUDAComment}{%
  \Comment{\makebox[1.85cm][c]{\colorbox{yellow!25}{\strut CUDA Core}}}%
}
\newcommand{\TensorComment}{%
  \Comment{\makebox[1.85cm][c]{\colorbox{green!25}{\strut Tensor Core}}}%
}
\newcommand{\FHEComment}{%
\Comment{\makebox[1.85cm][c]{\colorbox{lohitpurple!100}{\strut~~\textcolor{black}{FHECore~~}}}}%
}

%% file: src/0-abstract.tex
\begin{abstract}


Fully Homomorphic Encryption (FHE) enables computation directly on encrypted data but incurs massive computational and memory overheads, often exceeding plaintext execution by several orders of magnitude.
While custom ASIC accelerators can mitigate these costs, their long time-to-market and the rapid evolution of FHE algorithms threaten their long-term relevance.
GPUs, by contrast, offer scalability, programmability, and widespread availability, making them an attractive platform for FHE.
However, modern GPUs are increasingly specialized for machine learning workloads, emphasizing low-precision datatypes (e.g., INT$8$, FP$8$) that are fundamentally mismatched to the wide-precision modulo arithmetic required by FHE.
Essentially, while GPUs offer ample parallelism, their functional units, like Tensor Cores, are not suited for wide-integer modulo arithmetic required by FHE schemes such as CKKS. 
Despite this constraint, researchers have attempted to map FHE primitives on Tensor Cores by segmenting wide integers into low-precision (INT$8$) chunks. 
However, this incurs substantial overhead due to intermediate reassembly (for modulo reduction) and increased cross-core communication.

To overcome these bottlenecks, we propose \fhecore, a specialized functional unit integrated directly into the GPU's Streaming Multiprocessor. 
Our design is motivated by a key insight: the two dominant contributors to latency--Number Theoretic Transform and Base Conversion--can be formulated as modulo-linear transformations. 
This allows them to be mapped on a common hardware unit that natively supports wide-precision modulo-multiply-accumulate operations. 
\fhecore incorporates built-in Barrett reduction, eliminating the long sequences of instructions that would otherwise be required for modulo reduction; this consolidation strongly reduces the dynamic instruction count of applications.
Our simulations demonstrate that \fhecore reduces dynamic instruction count by a geometric mean of $2.41\times$ for CKKS primitives and $1.96\times$ for end-to-end workloads. 
These reductions translate to performance speedups of $1.57\times$ and $2.12\times$, respectively--including a $50\%$ reduction in bootstrapping latency--all while inuring a modest $2.4\%$ area overhead.

\end{abstract}

%% file: src/1-introduction.tex
\section{Introduction}
\label{sec:introduction}

Modern day computing increasingly relies on outsourced computations, where the servers gain full visibility of the client's plaintext data--an exposure unacceptable for applications handling sensitive information, such as financial records or medical diagnostics.
Fully Homomorphic Encryption (FHE) eliminates this vulnerability by enabling computation directly on encrypted data without ever exposing the underlying plaintext.
Over the years, researchers have successfully mapped FHE to a diverse range of application domains, including encrypted database queries~\cite{EDB1,EDB2}, private information retrieval~\cite{PIR1,PIR2}, encrypted image processing~\cite{IP1,IP2}, and, more recently, privacy-preserving machine learning~\cite{ellmo, PPML1,PPML2}.
While FHE paves the way for privacy-preserving computation, its practical adoption is hindered by severe computational and memory overheads.
Evaluating even simple functions on ciphertexts inflates execution times by $2$ to $5$ orders of magnitude as compared to their plaintext equivalents~\cite{OpenFHE, MemFHE}.
Therefore, realizing the potential of FHE necessitates the development of practical, FHE-capable computing systems.

FHE operations are highly parallel and can be formulated as SIMD-style computations.
This property naturally aligns them with general-purpose GPUs, which are capable of executing thousands of arithmetic operations concurrently.
Today, GPUs are deeply entrenched in cloud infrastructures, powering large-scale machine learning (ML) workloads and providing a mature execution environment.
Consequently, GPUs have emerged as the natural platform for accelerating server-side FHE computations, combining massive parallelism with a well-established programming ecosystem. 

A number of studies map FHE to ASIC designs \cite{f1, craterlake, bts, sharp, osiris, cinnamon, ARK}, which offer higher efficiency in terms of power, performance, and area than a GPU. 
Unfortunately, the cost of designing, verifying, and maintaining such hardware is prohibitively high, often requiring years of development and millions of dollars in investment~\cite{cifher, REED}. 
Likewise, FPGA-based FHE accelerators have made significant strides, however they continue to fall short of ASIC-level performance and are limited by small on-chip memories and low operating frequencies~\cite{FAB, HEAX, HEAP, book}.

Thus, while ASICs and FPGAs may deliver superior raw performance or energy efficiency, GPUs present the most accessible path toward deploying privacy-preserving applications at scale.
Realizing this potential, however, requires innovations in mapping FHE computations onto the GPU, thereby bridging the gap between FHE's algorithmic structure and the GPU's architectural capabilities.
To date, several efforts have addressed this challenge through optimized software libraries that fuse algorithmic advancements with high-performance programming techniques~\cite{FIDESlib, Cheddar,100x,PhantomFHE, HEonGPU}.

Over the past few years, GPUs have evolved to include specialized matrix-multiply units, such as NVIDIA's Tensor Cores and AMD's Matrix Core Engines, which deliver exceptional throughput for narrow-precision datatypes.
Recognizing this capability, recent works \cite{cross,tensorfhe,warpdrive, neo} have explored repurposing these `AI-oriented' cores for FHE computations.
However, all of them encounter a common limitation: current matrix units are optimized for floating-point or low-precision integer arithmetic, while FHE computations require wide-precision modulo arithmetic. 
The resulting need to decompose and reassemble operands introduces significant overhead--accounting for nearly $40\%$ of total Number Theoretic Transform (NTT) latency~\cite{warpdrive}. 
This makes a compelling case for hardware that natively caters to FHE-specific arithmetic.
Lacking native modulo-friendly opcodes forces FHE computations into long, fine-grained instruction chains, amplifying front-end pressure and overall latency. 
Such overheads further reinforce the need to consolidate many operations into a single, coarse-grained primitive.

These constraints raise a broader architectural question: \textit{How can we rearchitect the GPU to efficiently run FHE workloads without compromising performance of conventional high-performance applications?}
Conceptually, operations such as modulo arithmetic can be incorporated within existing GPU pipelines.
However, such modifications are intrusive in nature.
They may introduce additional pipeline stages, alter datapath widths and register formats, potentially degrading the efficiency of broader GPU workloads.
An alternate strategy is to introduce a dedicated functional unit specialized for modulo arithmetic that coexists alongside existing compute infrastructure.
This approach isolates FHE-specific computations whilst preserving the existing GPU microarchitecture.

The rest of this paper explores the viability of these directions and presents \fhecore, a specialized functional unit designed to accelerate FHE operations, without compromising general-purpose GPU performance.
To the best of our knowledge, GME~\cite{GME} is the only other work that introduces architectural changes specifically designed to accelerate FHE on GPUs.
In which, Shivdikar et al. extend AMD's MI$100$ GPU with an on-chip network, \mbox{$64$-bit} modulo-multiply-accumulate units, and a custom block scheduler.
While these enhancements deliver notable performance gains over the architectural baseline, they incur a $26.6\%$ area overhead--pushing the overall die size beyond present day lithographic reticle limits~\cite{bigchip}.

Effectively, executing FHE workloads on existing GPUs exposes two key limitations: $1)$ a mismatch in supported data types, and $2)$ inefficient realization of modulo arithmetic.
~To address these limitations, we propose \fhecore, a specialized functional unit integrated into the GPU architecture.
\fhecore mitigates existing datatype constraints and computational inefficiencies while maintaining a minimal silicon footprint--ensuring the design remains well within physical fabrication limits.



In this work, we make the following contributions:

\begin{itemize}
    \item An architectural enhancement -- \fhecore, integrated into the Streaming Multiprocessor (SM) to accelerate modulo-linear transformations central to FHE workloads, and engineered to incur minimal area overhead on the GPU while fully reusing its existing microarchitecture.

    \item An ISA extension that introduces a new instruction capable of performing modulo matrix multiplication. 
    This opcode consolidates long sequences of arithmetic and predication instructions into one operation, reducing the overall instruction count in workloads. 
    
    \item We demonstrate how Tensor Core-based NTT acceleration can benefit from using \fhecore. 
    Apart from CKKS primitives, we extend our evaluation to end-to-end workloads, including an encrypted language model, BERT-Tiny. 
\end{itemize}


We use a combination of tools in our evaluations to determine performance gains and estimate area overheads. 
We use NVBit~\cite{NVBit} and Accel-Sim~\cite{accelsim} to collect and simulate instruction traces of applications developed using FIDESlib~\cite{FIDESlib}. 
For latency and area estimation, we implement \fhecore in Verilog HDL and utilize SiliconCompiler~\cite{siliconcompiler} to perform the full physical design flow--including synthesis, placement, and routing--targeting the ASAP$7$~\cite{asap7} process node.

Our analysis shows geometric-mean speedups of $1.57\times$ for CKKS primitives and $2.12\times$ for full workloads, while increasing the A$100$ GPU die area by only $2.4\%$.
Across full workloads, the proposed ISA extension reduces dynamic instruction count by $2.41\times$, and by $1.96\times$ for CKKS primitives  and full workloads respectively.


%% file: src/2-background.tex
\section{Background}
\label{sec:background}
In this section, we analyze the computational characteristics of Fully Homomorphic Encryption (FHE), specifically focusing on the CKKS scheme and its primitives.
We then provide an overview of the \mbox{NVIDIA A$100$ GPU} and outline its programming model, which together form the architectural baseline for \fhecore{'s} integration.

\subsection{FHE Operations and Basics of CKKS}
\label{subsec:fhe-operations-and-basics-of-ckks}
\input{tables/notation}
\input{tables/primitives}
\input{figures/workloads}
Over the years, researchers have proposed several FHE schemes--BFV~\cite{BFV}, BGV~\cite{BGV}, CKKS~\cite{CKKS}, and TFHE~\cite{TFHE}--each suited to specific computation needs.  
Despite their differences, these schemes share a common computational backbone composed of kernels, like Number Theoretic Transform (NTT), Base Conversion, and element-wise modulo operations.  
When a plaintext application is mapped onto the encrypted domain, its computations are realized through compositions of these kernels.
As a result, the execution profile of any FHE program can be viewed as a repeated sequence of these basic kernels.  

In this work, we focus on the CKKS scheme, which enables approximate computation over fractional values. 
This makes it well-suited for ML workloads, where numerical errors are surprisingly tolerable~\cite{error}.  
CKKS leverages the Residue Number System (RNS) representation, enabling practical arithmetic on ciphertext polynomials.
Table~\ref{tab:notation} presents the notation used in CKKS-RNS, and Table~\ref{tab:primitives} summarizes its main primitives.


To understand the limitations of current GPUs on FHE workloads, we profiled four representative applications: Bootstrapping, Logistic Regression (LR) training, ResNet$20$, and \mbox{BERT-Tiny}.
After addressing the memory-boundedness of the workloads using established prior work~\cite{mad}, we note that forward and inverse NTTs dominate execution time, together accounting for $66\%$ of total runtime across these workloads.
The next-largest contributors are scalar operations, which account for $16.4\%$ of the runtime, followed by base conversion at $12.6\%$.
All remaining kernels--including automorphism--collectively contribute only $5\%$. 
Figure~\ref{fig:workloads} shows the exact breakdown of these kernels in our workloads.

The ensuing discussion examines NTT and Base Conversion in detail, showing that both can be reformulated as modulo-linear transformations; therefore, making them ideal candidates for acceleration on \fhecore.

\subsubsection{Number Theoretic Transform (NTT)}
NTT is the modulo-arithmetic analog of the Discrete Fourier Transform and is used to accelerate polynomial multiplications.  
NTT transforms a polynomial from its coefficient domain to its evaluation domain, enabling element-wise multiplication.  
For a vector $\mathbf{a}=[a_0, a_1, \dots, a_{N-1}]$ and a primitive $N$-th root of unity $\omega$ modulo $q_i$, the transform is defined as: 
\begin{equation}
    \hat{a}_k = \sum_{j=0}^{N-1} a_j \cdot \omega^{jk} \bmod q_i
\end{equation}
This operation corresponds to multiplying vector `$\mathbf{a}$' by an $N \times N$ (Vandermonde) matrix over $\mathbb{Z}_{q_i}$.

For practical security, CKKS typically requires ring dimensions in the range of $2^{15}$ to $2^{17}$~\cite{HEstandard}.  
In this scenario, rather than computing the full transform directly, software implementations adopt a hierarchical decomposition.
This was originally proposed by Bailey~\cite{bailey} to do Fast Fourier Transforms. 
The one-dimensional input vector is reshaped into a matrix of dimensions $N_1 \times N_2$, and the transform is computed as a collection of smaller NTTs along one dimension, followed by element-wise scaling and a second pass of NTTs along the other dimension: 
\begin{equation}
\hat{a}_{k_1 + k_2 N_1} =
\sum_{j_2 = 0}^{N_2 - 1}
\left(
\sum_{j_1 = 0}^{N_1 - 1}
a_{j_1 N_2 + j_2} \cdot \omega_{N_1}^{j_1 k_1}
\right)
\cdot \omega_{N}^{(k_1 + k_2) j_2}
\end{equation}
This decomposition restructures the computation as a hierarchical $4$-step NTT. 
Kim et al.~\cite{4step-ntt} demonstrate that this formulation exposes fine-grained parallelism and improves utilization of resources on GPUs.

\subsubsection{Base Conversion}
Base conversion is a fundamental operation in CKKS, used to switch between different RNS bases during rescaling, key switching, and bootstrapping.  
Let $Q = q_0\times q_1\times \dots\times q_{L-1}$ and $P = p_0\times p_1\times \dots\times p_{\alpha-1}$ denote two moduli and their corresponding composition for RNS.

The goal of base conversion is to generate a representation $\hat{a} \in \mathbb{Z}_{Q}$ from a polynomial $a \in \mathbb{Z}_{P}$. 
Let $a[n][m]$ and $\hat{a}[n][m]$ denote the $m^{\text{th}}$ residue of the $n^{\text{th}}$ coefficient of the polynomial $a$ and its transformed counterpart $\hat{a}$, respectively.
This is performed using the following relation, for $0\leq i < L$: 
\begin{equation}
\hat{a}\left[n\right][i] = 
\sum_{j=0}^{\alpha - 1} 
\left(
\left[ a\left[n\right][j] \cdot \hat{P}_j^{-1} \right]_{p_j}
\cdot
\hat{P}_j
\right)
\bmod q_i
\label{eq:baseconv}
\end{equation}
where $p_j$ is the $j$-th modulus in $P$, $P^{*} = \prod_{j=0}^{\alpha-1} p_j$, $\hat{P}_j = \frac{P^{*}}{p_j}$, and $[x]_p$ denotes reduction of $x$ modulo $p$.
Operationally, Equation~(\ref{eq:baseconv}) can be interpreted as a dot product, which can be extended to a modulo matrix--matrix multiplication when converting all coefficients, $n\in\left[0,N-1 \right]$ of the polynomial (more on this in \S~\ref{subsec:mapping-base-conversion}).

\subsection{NVIDIA A100 GPU Architecture}
\label{subsec:nvidia-a100-gpu-architecture}
\input{figures/tensor-core-programmability}

Next, we outline the architectural features and programming model of the NVIDIA A$100$ GPU, which serves as a representative example of a modern data center accelerator.  
This overview provides the necessary context for analyzing the system-level effects of integrating a specialized functional unit into an existing GPU pipeline.
Although the discussion centers on A$100$, the \fhecore design is in no way tied to this GPU and can be readily extended to newer ones.

The A$100$ GPU system consists of $108$ Streaming Multiprocessors (SMs) integrated within a single die. 
Each SM contains multiple execution units, including CUDA cores and Tensor Cores, as well as load/store units, register files, and shared memory, enabling massive thread-level parallelism. 
Threads are grouped into warps, which execute in a SIMT fashion. 
At the system level, the GPU employs a hierarchical memory architecture comprising $40$ or $80$~GB of off-chip memory, $40$~MB of L$2$ cache shared across all SMs, $192$~KB of L$1$/shared memory, and a $256$KB register file within each SM. 
The total die area of the A$100$ GPU is reported as $826$~mm$^2$~\cite{nvidia_ampere}.

Tensor Cores have been part of NVIDIA GPUs since the Volta architecture~\cite{nvidia_volta} and have continued to evolve over subsequent generations. 
In a typical execution pipeline, threads issue load/store instructions to fetch data from the memory hierarchy into the register file, from which the Tensor Core reads inputs and writes results. 
In the Volta and Ampere architectures, Tensor Cores themselves do not interact directly with the memory hierarchy, but instead operate exclusively on values stored in registers.
A key enhancement introduced in Ampere is the ability of Tensor Cores to execute concurrently with integer and floating-point units. 
Raihan et al.~\cite{modeling-tensor-cores} argue that this is possible due to the separation of register file access paths. 
Unlike prior generations, where CUDA and Tensor Cores competed for shared ports, Ampere assigns distinct access ports to each, reducing contention and enabling greater concurrency.

The A$100$ GPU contains $432$ Tensor Cores, uniformly distributed at four per SM.  
The peak throughput varies with operand precision, ranging from $9.7$~TOPS for FP$64$ to $1248$~TOPS for INT$4$, at a peak frequency of $1410$~MHz~\cite{nvidia_ampere}.  
Although the exact microarchitecture of Tensor Cores is not publicly disclosed, prior studies have attempted to reverse engineer their details~\cite{modeling-tensor-cores, Modeling-tensor-cores-2, Modeling-tensor-cores-3}.  
These works collectively uncover key traits, including the arrangement of tile fragments in registers, the cycle-level latencies of MMA instructions, and the interface to the register file.

Figure~\ref{fig:tensor-core-programmability} illustrates how the Tensor Core execution model is exposed through a layered compilation framework spanning high-level CUDA APIs to machine instructions.  
At the software level, CUDA's \texttt{wmma} API enables matrix operations on fixed-size ($16\times16$) tiles and supports multiple operand types, as summarized in Table~\ref{tab:precisions}.  
These MMA functions are lowered by the compiler to PTX~\cite{PTX} instructions, such as \texttt{wmma.load}, \texttt{wmma.mma.sync}, and \texttt{wmma.store}, which abstract away hardware details.  
Finally, PTX instructions are compiled into SASS micro-operations, such as \texttt{IMMA}, which activate Tensor Cores.  

Tensor Cores execute only matrix-multiply operations - MMA (green-shaded region), while the others (beige-shaded region) are handled by load/store (LD/ST) units to move data between memory and the register file.  
In the SASS instruction stream, the identifier following `\texttt{IMMA}' (i.e. \texttt{16816}) encodes the dimensions of the matrix operation using the $m\times n\times k$ notation.  
For a systolic-array implementation, this identifier would represent the physical dimensions of the array (e.g., $16\times8$ or $8\times16$).

%% file: tables/notation.tex
\begin{table}
\centering
\caption{Notation used in the CKKS-RNS scheme.}
\vspace{-5pt}
\label{tab:notation}
\resizebox{\columnwidth}{!}{
\begin{tabular}{ll}
    \toprule
    \textbf{Symbol} & \textbf{Description} \\
    \midrule
    $N$ & Polynomial ring dimension. \\[2pt]
    $L$ & Multiplicative depth. \\[2pt]
    $Q = \{ q_0, \dots, q_L \}$ & RNS moduli chain. \\[2pt]
    $R_Q = \prod_{i=0}^{L} R_{q_i}$ & Polynomial ring for CKKS-RNS. \\[2pt]
    $c = (c_0, c_1) \in (R_Q)^2$ & Ciphertext with two polynomials in $R_Q$. \\[2pt]
    $\alpha$ & Number of extended moduli. \\[2pt]
    $P = \{ p_0, p_1, \dots, p_{\alpha-1} \}$ & Extended moduli chain for key switching. \\[2pt]
    $\omega_N$ & Primitive $N$-th root of unity. \\
    \bottomrule
\end{tabular}
}
\vspace{-5pt}
\end{table}

%% file: tables/primitives.tex
\begin{table*}
\centering
\caption{Summary of CKKS primitive operations and their corresponding mathematical transformations.}
\vspace{-5pt}
\resizebox{\textwidth}{!}{
\begin{tabular}{p{2.3cm} p{6.7cm} p{7.5cm}}
 \toprule
 \textbf{Primitive} & \textbf{Output} & \textbf{Description}  \\
 \midrule
 PtAdd($c, p$) & $(c_0 + p, c_1)$ & Adds a plaintext polynomial to a ciphertext polynomial. \\[2pt]
 HEAdd($c, c'$) & $(c_0 + c'_0, c_1 + c'_1)$ & Adds two ciphertexts coefficient-wise. \\[2pt]
 PtMult($c, p$) & $(\lfloor p \cdot c_0 / q_i \rceil, \lfloor p \cdot c_1 / q_i \rceil)$ & Multiplies a plaintext with a ciphertext and rescales. \\[2pt]
 KeySwitch($c_s, \text{evk}_{s'}$) & $(c_0, c_1 \cdot \text{evk}_{s'})$ & Switches the ciphertext to a new secret key. \\[2pt]
 Rescale($c, q_i$) & $(\lfloor c_0/q_i \rceil, \lfloor c_1/q_i \rceil)$ & Divides both ciphertext polynomials by the scale $q_i$. \\[2pt]
 HEMult($c, c', \text{evk}_{s^2}$) & $\text{Rescale}((c_0 c'_0, c_0 c'_1 + c_1 c'_0) + \text{KeySwitch}(c_1 c'_1, \text{evk}_{s^2}))$ & Multiplies two ciphertexts, then performs KeySwitch and Rescale. \\[2pt]
 Rotate($c, k$) & $(\Phi(c_0), 0) + \text{KeySwitch}(\Phi(c_1), \text{evk}_{\Phi(s)})$ & Rotates the encrypted vector by $k$ slots. Here `$\Phi$' denotes a slot-wise rearrangement step known as automorphism. \\
 \bottomrule
\end{tabular}
}
\label{tab:primitives}
\vspace{-10pt}
\end{table*}

%% file: figures/workloads.tex
\begin{figure}
    \centering
    \includegraphics[width=\linewidth]{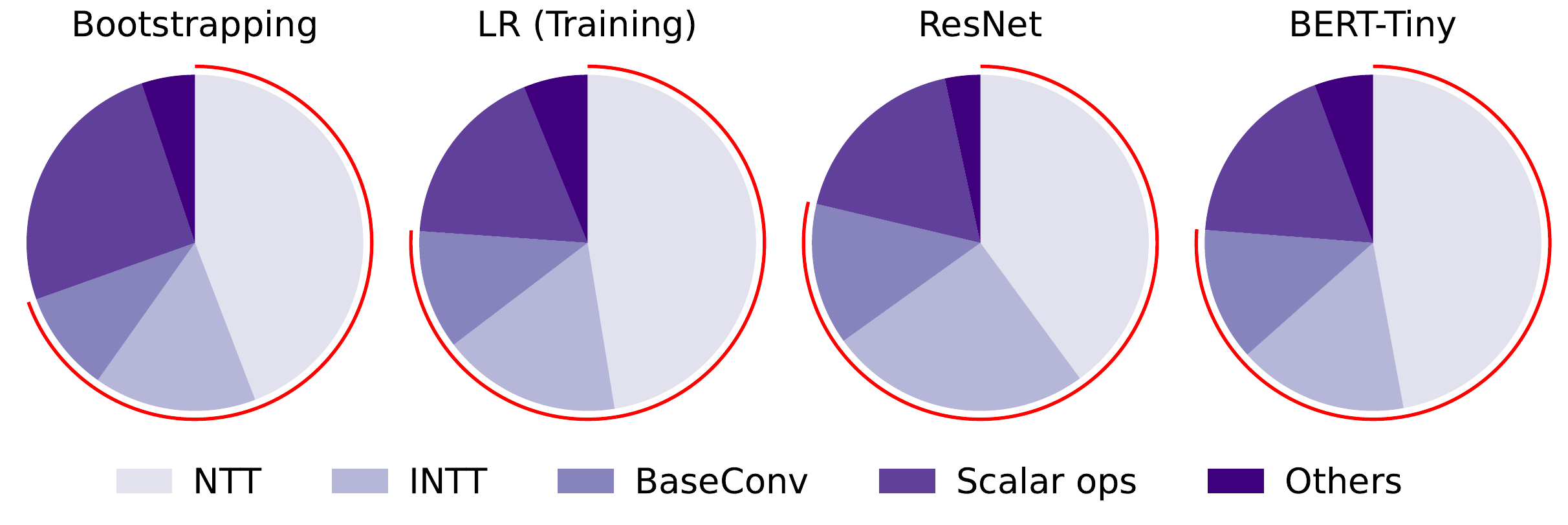}
    \vspace{-25pt}
    \caption{Latency decomposition of CKKS-based workloads on the A$100$ GPU. Here FIDESlib resolves the memory boundedness of CKKS workloads using established prior work~\cite{mad}, and then we shift our focus on the compute performance. Across bootstrapping, logistic-regression (LR), ResNet$20$, and BERT-Tiny; the NTT, INTT, and BaseConv steps together account for more than $70 \%$ of total runtime, identifying them as the dominant compute performance bottlenecks.}
    \label{fig:workloads}
    \vspace{-12pt}
\end{figure}

%% file: figures/tensor-core-programmability.tex
\begin{figure*}
    \centering
    \includegraphics[width=\linewidth]{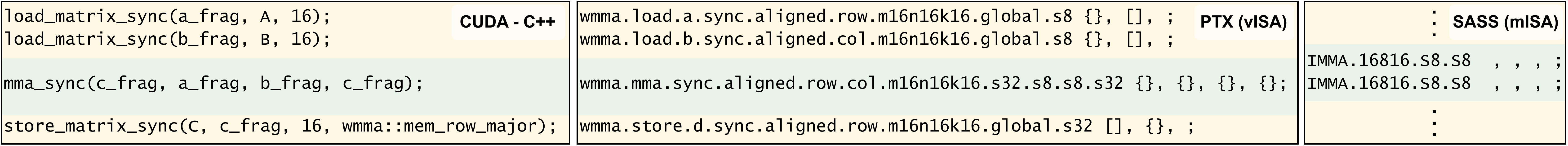}
    \vspace{-20pt}
    \caption{Compilation flow for the Ampere architecture Tensor Core's INT$8$ operations. 
    High-level CUDA C++ APIs using \texttt{mma} intrinsics are first lowered to PTX (vISA) instructions, which are further lowered to SASS (mISA) instructions. 
    The green highlighted sections sandwiched in the middle represent instructions that are executed by Tensor Cores, while the remaining instructions are handled by the load/store (LD/ST) units to move data within the register file and across the hierarchy.}
    \label{fig:tensor-core-programmability}
    \vspace{-10pt}
    \vspace{-2pt}
\end{figure*}

%% file: src/3-motivation.tex
\section{Motivation}
\label{sec:motivation}
\input{tables/precisions}

In this section, we outline the system/architectural limitations that prevent GPUs from efficiently executing FHE workloads and motivate the need for dedicated hardware support.

\subsubsection{Growing Mismatch Between GPU Datatypes and FHE Requirements}
Today's GPUs increasingly prioritize throughput on low-precision datatypes such as INT$8$, FP$8$, and INT$4$~\cite{nvidia_blackwell}.  
As Table~\ref{tab:precisions} shows, the trend to support narrow-datatypes on tensor cores has steadily grown, while native INT$32$ support in CUDA cores has been phased out in the Blackwell architecture~\cite{blackwell_int32}.  
This evolution is well-aligned with the needs of ML workloads, but it stands in stark contrast to the needs of FHE, whose arithmetic is dominated by wide-precision integer operations (INT$32$ or INT$64$).
As a result, the numerical requirements of FHE increasingly diverge from the datatype capabilities prioritized in modern GPU designs.

\subsubsection{The Need for Coarser-Grained Execution in FHE}
Modern GPUs expend substantial energy not only on arithmetic but on the front-end instruction processing driven by hundreds of Streaming Multiprocessors (SMs).
Each SM independently performs instruction fetch, decode, issue, and scheduling for every instruction in the program.
Tensor cores mitigate this overhead by raising the granularity of execution: a single Tensor Core instruction triggers an entire MMA operation that would otherwise require hundreds of scalar machine instructions. 
Collapsing many operations into one instruction substantially reduces the front-end activity of SMs.


FHE workloads, however, fail to fully realize these benefits. 
Even when coarse-grained Tensor Core operations are available, in FHE each arithmetic step must be followed by a modulo-reduction step, generating long chains of CUDA core instructions. 
This fine-grained instruction stream drives up front-end work and, in turn, energy consumption.
This limitation motivates a coarse-grained unit for FHE--analogous to Tensor Cores but specialized for wide-precision modulo arithmetic. 
By performing multiply-accumulate and modulo reduction within a single instruction, such a unit would sharply reduce dynamic instruction count and alleviate front-end energy overheads.


\subsubsection{GPUs provide the most realistic path for FHE acceleration}
Despite their increasing specialization for AI workloads, modern GPUs retain several architectural strengths that make them the most viable platform for large-scale FHE deployment.

\begin{enumerate}[label=\roman*.]
    \item\textit{Scalability:} Multi-GPU scaling is already a first-class design goal in deep-learning systems.
    Interfaces such as NVLink~\cite{nvidia_nvlink} provide the high-bandwidth, low-latency communication required for distributed FHE, as demonstrated in recent multi-GPU FHE works~\cite{cufhe, multigpufhe}.

    \item\textit{Advanced memory systems:} GPUs incorporate sophisticated memory systems well aligned with FHE’s data-movement demands.  
    The Tensor Memory Accelerator enables asynchronous overlapped transfers~\cite{nvidia_blackwell}, while Hopper’s distributed shared memory improves locality for bandwidth-bound kernels~\cite{nvidia_hopper}.  
    Notably, GME~\cite{GME} reports its largest gains from an inter-SM network-on-chip--an architectural feature now natively part of Hopper GPUs (Distributed Shared Memory).
    
    \item\textit{``The CUDA moat'':} The CUDA programming stack provides a mature software development ecosystem.  
    GPU-based extensions integrate cleanly into this ecosystem, unlike ASIC or FPGA solutions that require separate compiler infrastructure and software stacks.  
    This ecosystem also enables rapid prototyping of emerging FHE algorithms.
\end{enumerate}
Taken together, these factors make GPUs the most realistic deployment vehicle for high-performance encrypted computation.
This work builds on that premise by introducing \fhecore, a GPU-integrated functional unit purpose-built to execute wide-precision modulo-linear transformations central to FHE workloads.

%% file: tables/precisions.tex
\begin{table}[t]
\centering
\caption{Data Precision Support Across GPU Architectures. \\ $\text{\#TC = Number of Tensor Cores, and}$ \\  \#SM = Number of Streaming Multiprocessors.}
\vspace{-5pt}
\label{tab:precisions}
\resizebox{\columnwidth}{!}{
{\scriptsize
\setlength{\tabcolsep}{3.5pt}
\begin{tabular}{
    ccc                      
    ccccccccc                
    ccc>{\columncolor{highlightcol}}cc 
}
\toprule
\multirow{2}{*}{\textbf{GPU}} & \multirow{2}{*}{\#TC} & \multirow{2}{*}{\#SM} 
& \multicolumn{9}{c}{\textbf{Tensor Cores}} & \multicolumn{5}{c}{\textbf{CUDA Cores}} \\
\cmidrule(lr){4-12} \cmidrule(lr){13-17}
& & 
& \rot{FP$64$} & \rot{TF$32$} & \rot{FP$16$} & \rot{BF$16$} & \rot{FP$8$} & \rot{FP$6$} & \rot{INT$8$} & \rot{INT$4$} & \rot{INT$1$}
& \rot{FP$32$} & \rot{FP$16$} & \rot{BF$16$} & \rot{INT$32$} & \rot{INT$8$} \\
\midrule
\makecell{V$100$\\(Volta)}         & $640$ & $80$  & -- & -- & \cmark & -- & -- & -- & -- & -- & -- & \cmark & \cmark & -- & \cmark & \cmark \\[6pt]
\makecell{RTX$6000$\\(Turing)}     & $576$ & $72$  & -- & -- & \cmark & -- & -- & -- & \cmark & \cmark & \cmark & \cmark & \cmark & -- & \cmark & \cmark \\[6pt]
\makecell{A$100$\\(Ampere)}        & $432$ & $108$ & \cmark & \cmark & \cmark & \cmark & -- & -- & \cmark & \cmark & \cmark & \cmark & \cmark & \cmark & \cmark & \cmark \\[6pt]
\makecell{H$100$\\(Hopper)}        & $528$ & $132$ & \cmark & \cmark & \cmark & \cmark & \cmark & -- & \cmark & \cmark & \cmark & \cmark & \cmark & \cmark & \cmark & \cmark \\[6pt]
\makecell{B$100$\\(Blackwell)}     & $528$ & $132$ & \cmark & \cmark & \cmark & \cmark & \cmark & \cmark & \cmark & \cmark & \cmark & \cmark & \cmark & \cmark & \red{--} & \red{--} \\
\bottomrule
\end{tabular}
}
}
\vspace{-10pt}
\end{table}

%% file: src/4-fhecore.tex
\section{\fhecore}
\label{sec:fhecore}
\input{figures/microarch}

In this section, we describe the integration of \fhecore within the NVIDIA A$100$ GPU.
The discussion follows a top-down system organization, detailing the architectural design and the rationale behind key design choices.  
Subsequently, we analyze the implications of dataflow for modulo matrix-multiplication when executed on \fhecore.  
We then outline the extensions at various levels of the CUDA programming framework that enable the use of the new functional unit.
Finally, we conclude by discussing the implications of extending Tensor Cores to support modulo arithmetic.

\subsection{Overall System Architecture}
\label{subsec:system-architecture}
On the left of Figure~\ref{fig:microarch}, we show a high-level view of our system's architecture.
On the right, we show how each Streaming Multiprocessor (SM) incorporates \fhecore units alongside existing CUDA cores and Tensor Cores.  
In this configuration, every SM hosts an equal number of \fhecore and Tensor Cores, which share the register file and warp scheduler with other functional units.  

Such an arrangement, with a tight coupling of general-purpose compute cores and \fhecore{s}, supports data exchange through the SM's register file.  
This is also in line with the existing GPU design, where compute cores maintain a clear boundary separating them from the rest of the system.
This allows \fhecore to fully reuse the existing memory hierarchy, leveraging features such as asynchronous data transfers, shared memory, and on-chip interconnects.

In this work, we use the NVIDIA A$100$ GPU as a representative platform. This choice is solely driven by the availability of ``tested'' A$100$ configuration files in \mbox{Accel-Sim}~\cite{accelsim}, as well as our access to the real A$100$ system for trace collection.  
While the details in this section target an A$100$, the \fhecore design can be readily deployed in newer architectures, and the broader conclusions of this work would remain applicable.

\subsection{Integration of \fhecore}
\label{subsec:integration-of-fhecore}
\fhecore is designed to accelerate modulo-linear transformations, which in essence can be viewed as tensor operations.  
Modern GPUs already include dedicated Tensor Cores optimized for dense -- and, in recent architectures, structured sparse -- matrix multiplications in ML workloads.  
Inspired by this model, we integrate \fhecore as an independent functional unit within each SM, thereby providing a clean, `modulo' capable extension to the existing compute fabric.

Deploying \fhecore as a standalone unit opens up interesting design decisions to be made.  
First, we instantiate the exact same number of \fhecore units as Tensor Cores per SM.  
This symmetry ensures compatibility with the existing warp dispatch unit and allows reuse of the SASS instruction scheduling pattern that governs tensor-core execution.  
As a result, no modifications are required in either the compiler stack or the instruction stream sequence to support \fhecore execution - more details on this are discussed in \S~\ref{subsec:programmability-and-isa-extension}.  

Second, this configuration simplifies access to the register file, illustrated in the SM diagram in Figure~\ref{fig:microarch}.
Given that the register ports and datapath already support $32$-bit floating-point operands for Tensor Cores, they can naturally accommodate the $32$-bit integer operands used by \fhecore, requiring no modifications to the bus interface or register-file design.  

Finally, to minimize architectural overhead, we propose that \fhecore and Tensor Cores share common read and write ports to the register file.  
This decision is guided by the findings of Raihan et al.~\cite{modeling-tensor-cores}, who observe that simultaneous execution is not possible when register ports are shared across functional units.  

In practice, FHE and plaintext ML workloads are mutually exclusive.  
Once an application is encrypted, all computations operate on ciphertexts using modulo arithmetic, leaving floating-point and mixed-precision units in the Tensor Core idle.  
Therefore, no realistic workload would require concurrent activation of both Tensor Cores and \fhecore{s} within the same SM.  
Given this separation, port sharing provides an area-efficient integration strategy without affecting the performance of the FHE and non-FHE workloads on GPUs.

\subsection{Microarchitecture}
\label{subsec:microarchitecture}
As shown in Figure~\ref{fig:microarch}, \fhecore is implemented as a two-dimensional systolic array of Processing Elements (PEs) that are capable of performing modulo-linear transformations.  
In \S~\ref{subsec:fhe-operations-and-basics-of-ckks} we discussed that among the core arithmetic kernels, the forward and inverse NTT together account for $66\%$ of the overall runtime in encrypted workloads (see Figure~\ref{fig:workloads}), making them prime candidates for acceleration.  
While a dedicated accelerator tailored exclusively for NTT will offer better performance, such specialization would limit applicability to other primitives.  
Instead, \fhecore adopts a systolic array structure to balance generality and performance--enabling efficient acceleration of both (I)NTT and base-conversion in FHE.
This microarchitecture also makes the functional unit suitable for similar modulo-linear transformations in other cryptographic algorithms, such as Zero-Knowledge Proofs and Post-Quantum Cryptography.

The systolic array is organized as a $16\times8$ grid of PEs, capable of performing matrix multiplications of the form $16\times8\times16$, mirroring the behavior of Tensor Cores in the A$100$ GPU.  
This design choice is pragmatic: aligning with tensor-core operand configurations avoids speculative assumptions about proprietary internals of the GPU, such as the number of register-file ports or the bandwidth supported by the buses.  
It also ensures that existing buses and port widths can sustain the required data movement for \fhecore operations without modifying any of the SM's internals.  

Each PE of the systolic array computes $\red{a}\cdot \blue{b} \bmod q$ over $32$-bit operands.
This computation can be efficiently realized by including a multiplier and Barrett reduction module within each PE~\cite{Sapphire}.  
The hardware implementation of Barrett reduction eliminates long instruction sequences, which comprise of multiple add, multiply, and predicate operations, thus substantially reducing the dynamic instruction count of the program.  
The PE is internally pipelined with six stages, producing one result per cycle--in the output-stationary dataflow (discussed in \S~\ref{subsec:dataflow}). 
This ensures full utilization across the array during the computation.

We use Barrett reduction as the modulo reduction technique within \fhecore because it is broadly applicable across primes and avoids the additional pre- and post-processing required by other techniques (like  Montgomery~\cite{montgomery} and Shoup~\cite{shoup}).
In practice, the choice of reduction method is not user-selectable, and software libraries such as OpenFHE automatically determine the appropriate technique based on operand characteristics~\cite{OpenFHE}.
Consequently, tying \fhecore to Barrett reduction does not restrict or alter any user-applications.
The only modification required would be at the library layer, where maintainers would update their backend to use the new programming extensions that map directly to \fhecore operations.
Details on the programming extensions required for \fhecore are discussed in \S~\ref{subsec:programmability-and-isa-extension}.

\subsection{Dataflow Analysis}
\label{subsec:dataflow}

Systolic arrays, widely used in ML accelerators, use multiple dataflow patterns to maximize data reuse and minimize communication overhead~\cite{ML_SA1,ML_SA2,ML_SA3,ML_SA4}.  
The most common dataflows are input-stationary, weight-stationary, and output-stationary.  
In input- and weight-stationary modes, which we collectively refer to as operand-stationary dataflows, one operand matrix is programmed (stored and kept) within the PEs while the other is streamed across the array.  
Partial sums or intermediate results are propagated to neighboring PEs as computation proceeds.  
In the output-stationary dataflow, both operand matrices stream through the array, and each PE locally accumulates the partial products.  
The final outputs are written to memory only after the entire computation completes, thereby reducing register-file accesses.

For \fhecore, we adopt the output-stationary dataflow for modulo matrix multiplication.  
This decision is driven by the internal pipeline latency of each PE, which requires $6$ cycles to complete a single modulo multiply-accumulate operation of the form $R \leftarrow (R + \red{a} \cdot \blue{b}) \bmod q$. 
The $6$-cycle depth stems from the timing constraint required for \fhecore to operate faster than the GPU clock frequency.
Meeting this constraint ensures that \fhecore does not become part of the GPU chip design's critical path.

Figure~\ref{fig:dataflow} contrasts operand-stationary and output-stationary dataflows in a \fhecore systolic array.
In operand-stationary flow, one operand must traverse the full six-stage PE pipeline before propagating vertically, creating multi-cycle stalls and delaying downstream PE activation.
Output-stationary flow instead forwards both operands every cycle, preserving continuous data movement and eliminating pipeline bubbles.
As a result, output-stationary dataflow completes modulo-matrix multiplication in substantially fewer cycles and is the more efficient choice for \fhecore.

Now, based on the PE’s internal latency of $T = 6$ cycles, the total number of cycles required to perform a modulo matrix multiplication using a systolic array with $S_{R}$ rows and $S_{C}$ columns is given by $2S_{R} + S_{C} + T - 2$ cycles~\cite{scalesim}.  
Accordingly, \fhecore -- configured as a $16 \times 8$ systolic array -- can compute a $16 \times 8 \times 16$ matrix multiplication in $44$ cycles.
\input{figures/df_and_memsys}

\subsection{Interface with Memory Subsystem}
\label{subsec:memory-interface}

The addition of \fhecore introduces no modifications to the existing GPU memory hierarchy.
Similar to the A$100$ Tensor Cores, \fhecore{s} operate exclusively with the register file, fetching operands through register ports and writing results back, without directly interacting with shared memory or cache levels.
This decoupled design philosophy of the GPU is illustrated in Figure~\ref{fig:memory-interface}, wherein compute cores and the memory subsystem operate asynchronously, enabling seamless integration of \fhecore{s} within each SM.

Such integration preserves full compatibility with the existing CUDA programming framework, which already provides intrinsics to support data movement for compute kernels.
Consequently, existing memory-centric optimizations, such as the `asynchronous copy' mechanism introduced in the A$100$, which overlaps data transfers with warp execution, can be leveraged when invoking \fhecore operations.

\subsection{Programmability and ISA Extension}
\label{subsec:programmability-and-isa-extension}

We adopt the WMMA namespace structure (used to operate Tensor Cores) with a key modification to the computation step: we propose a new CUDA intrinsic called \texttt{fhe\_sync} (see Figure~\ref{fig:fhec}). 
This intrinsic operates identically to \texttt{mma\_sync}, but with support for modulo arithmetic by accepting additional arguments, namely the modulus $q$ and the precomputed Barrett constant $\mu$, which would be programmed into the PEs of \fhecore. 

The WMMA namespace also offers intrinsics to arrange matrix fragments in specific registers according to a predefined tiling layout, setting the stage for dataflow orchestration to compute in the functional unit~\cite{modeling-tensor-cores}. 
By reusing these intrinsics, \fhecore~inherits the register allocation and operand routing mechanisms, enabling seamless integration with minimal changes to the CUDA front-end and PTX intermediate representation.

CUDA intrinsics such as \texttt{fhe\_sync} can be lowered to PTX, thanks to NVIDIA’s documentation and support for inline PTX~\cite{PTX}. 
The final compilation step, which generates machine-level SASS instructions, is handled by the nvcc compiler backend; its implementation is not publicly accessible. 
As a result, we cannot directly introduce new SASS instructions from the PTX. To overcome this, we simulate the inclusion of \fhecore instructions by manually inserting them into the instruction trace. 
Its template mirrors the structure of Tensor Core SASS instructions (e.g., \texttt{IMMA.16816}) but are renamed with a new opcode (e.g., \texttt{FHEC.16816}) and register specifiers appropriate for 32-bit modulo arithmetic.
This manual insertion strategy allows us to evaluate the effect of \texttt{FHEC} on instruction count, and prototype \fhecore's integration without requiring access to the closed-source compiler stack.

In addition to improving programmability, introducing a dedicated instruction for modulo matrix multiplication provides a critical advantage in reducing dynamic instruction count. This is especially important for GPUs, which are widely deployed in data centers. By consolidating a collection of arithmetic and data-movement instructions into a single instruction, we reduce the front-end pipeline activity of each SM. 

\input{figures/fhec}

\subsection{Enhanced Tensor Cores for FHE}
\label{subsec:analysis-on-extending-tensor-cores}
An alternative approach to accelerating modulo-linear transformations is to extend existing Tensor Cores for FHE computations.
At first glance, this may seem more appealing than introducing a separate functional unit.
However, accurately gauging the feasibility of such an extension is non-trivial, primarily because NVIDIA does not disclose the internal microarchitecture of Tensor Cores.
Without access to details, such as the datapath organization or register-file port widths, any analysis of the required modifications would be highly speculative.

Despite these limitations, we attempt to estimate the area overhead that would be incurred when enhancing Tensor Cores.
We consider a plausible design abstraction of Tensor Cores modeled as a systolic array composed of arithmetic units that support the datatypes listed in Table~\ref{tab:precisions}. 
Within this abstraction, the proposed enhancement involves augmenting each PE with an additional arithmetic unit that supports $32$-bit integer multiply-accumulate operations (functionally absent in Tensor Cores), followed by Barrett reduction. 
This addition is sufficient to accelerate the modulo-linear transformations.

The area metrics for a $16\times8$ systolic array of Tensor Core PEs are reported in Table~\ref{tab:tensorcoreextension}.
Each PE incorporates FP$64/32/16$ and INT$8$ ALUs, matching the A$100$ Tensor Core design, and the values were obtained from RTL synthesis using SiliconCompiler on ASAP$7$.
With $432$ Tensor Cores on the A$100$, the total Tensor Core footprint is estimated to be $32.65~\text{mm}^{2}$.
When these are enhanced to accommodate support for modulo arithmetic, the area grows to $50.01~\text{mm}^{2}$, making the GPU area $843.36~\text{mm}^{2}$, which incurs a $2.1\%$ overhead and is within the reticle limit.

As discussed in \S~\ref{subsec:dataflow}, \fhecore~can compute a matrix multiplication in $44$ cycles. 
However, if one were to extend the capabilities of the Tensor Core, it would also inherit its constraints (like the critical path).
In Accel-Sim~\cite{accelsim}, Tensor Core instructions are modeled with a fixed latency of $64$ cycles, based on microbenchmarking results from Raihan et al.~\cite{modeling-tensor-cores}.
Thus, \fhecore must conform to this increased latency despite having the same modulo arithmetic capabilities.

\input{tables/tensorcoreextension}

%% file: figures/microarch.tex
\begin{figure*}
    \centering
    \includegraphics[width=\linewidth]{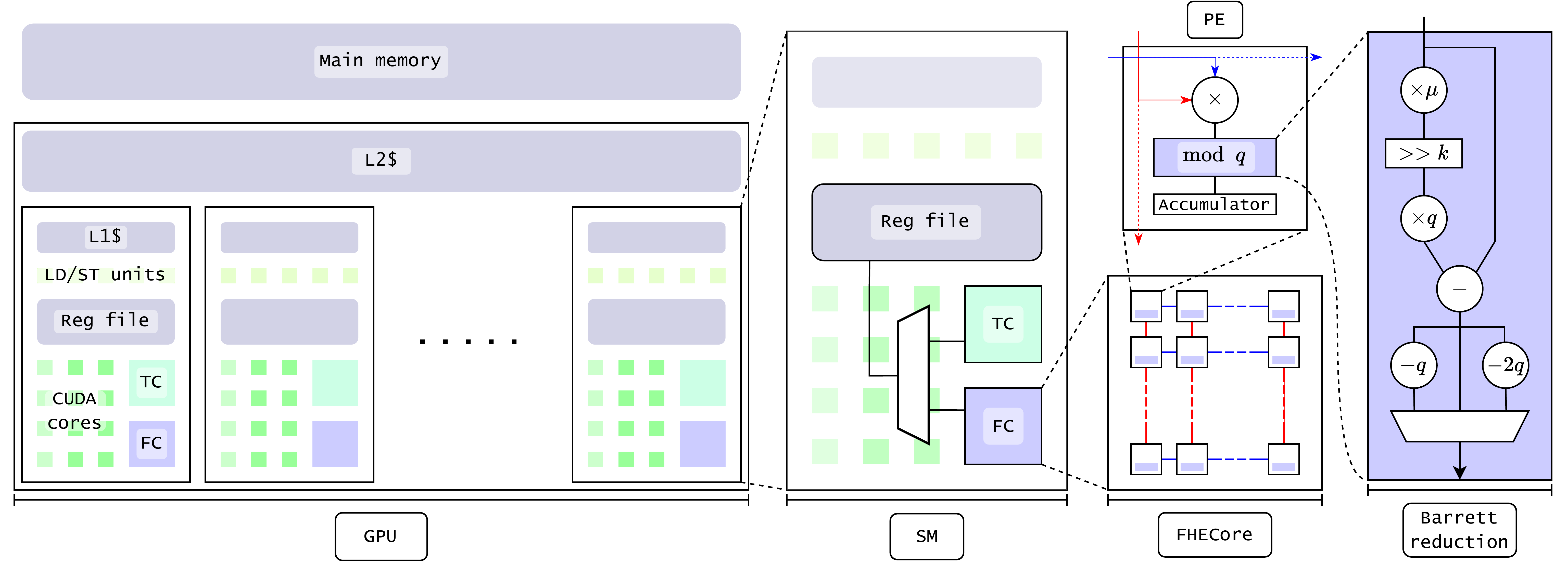}
    \vspace{-20pt}
    \caption{ \fhecore (FC) is introduced as a functional unit alongside existing CUDA cores and tensor cores (TC) within each Streaming Multiprocessor. 
    Like TC, it operates on values stored in the register file and does not directly interact with the caches or the main memory.
    The \fhecore consists of a $2$D systolic array of processing elements (PEs), each supporting modulo multiply-and-accumulate operations. 
    The PE depicted in this figure is to support the output-stationary dataflow, where partial products are accumulated within an internal register.
    The modulo reduction step is handled by a dedicated Barrett reduction pipeline.
    Here $\mu$ is the precomputed constant, and q is the modulus.}
    \label{fig:microarch}
    \vspace{-10pt}
\end{figure*}

%% file: figures/df_and_memsys.tex
\begin{figure*}
  \centering
  \begin{minipage}{0.61\linewidth}
    \centering
    \includegraphics[width=\linewidth]{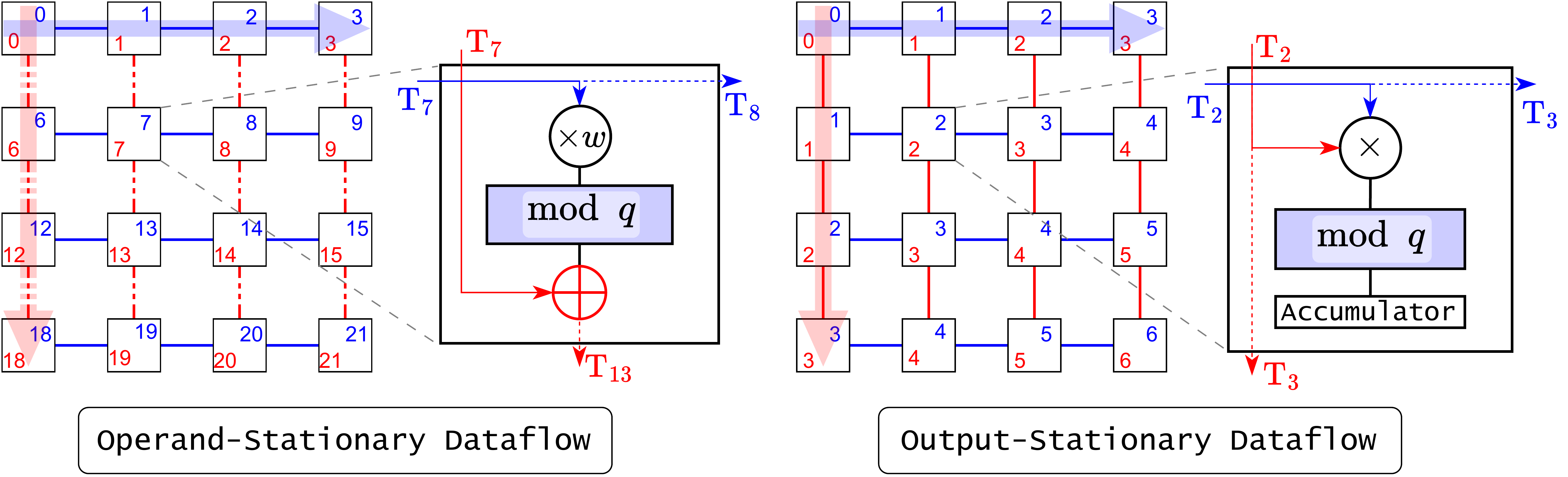}
    \vspace{-20pt}
    \caption{Traversal of data in operand-stationary vs. output-stationary dataflows on a miniaturized example $4\times4$ systolic array of \fhecore PEs.
    Here, the \red{red} and \blue{blue} inputs of each PE correspond to either entries or partial-sums of the matrices.
Numbers inside each PE denote the cycle in which the PE receives \red{red} and \blue{blue} operands for the first time.
In the operand-stationary dataflow, only the \blue{blue} operand advances each cycle, while the \red{red} operand must traverse the entire $6$-stage PE pipeline before forwarding a partial sum to the PE vertically below.
In contrast, the output-stationary dataflow forwards both operands (\red{red} and \blue{blue}) every cycle, allowing for a significantly faster modulo matrix multiplication.}
\vspace{-7pt}
    \label{fig:dataflow}
  \end{minipage}\hfill
  \begin{minipage}{0.37\linewidth}
    \centering
    \includegraphics[width=\linewidth]{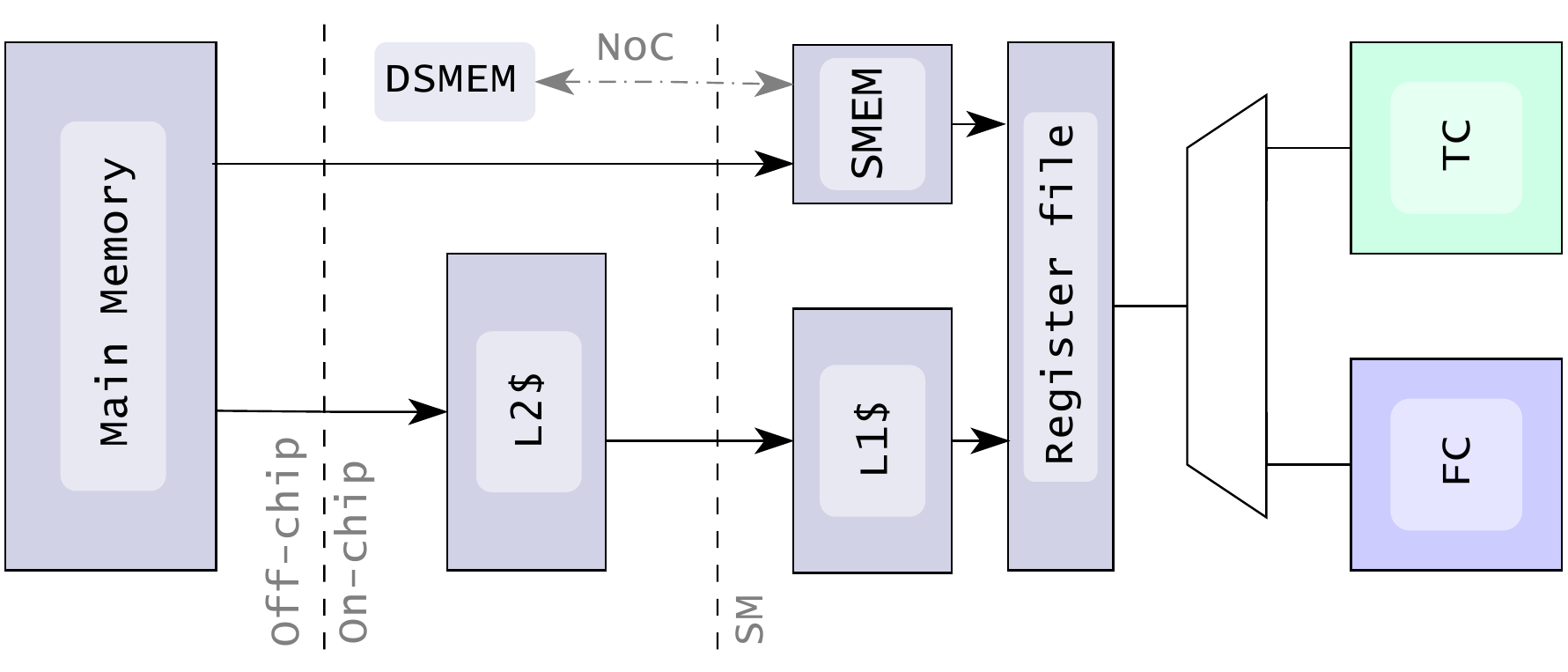}
    \vspace{-20pt}
    \caption{Interface of \fhecore with the GPU memory hierarchy.  
\fhecore (FC) shares the same register file and memory hierarchy as CUDA and Tensor Cores (TC).
Like Tensor Cores, \fhecore interacts solely with the register file, fetching operands and writing results via register ports.  
This design also allows future memory-system enhancements, such as distributed shared memory (DSMEM) introduced in the Hopper architecture, to be utilized by \fhecore.}
    \label{fig:memory-interface}
    \vspace{-7pt}
  \end{minipage}
  \vspace{-7pt}
\end{figure*}


%% file: figures/fhec.tex
\begin{figure}
    \centering
    \includegraphics[width=\linewidth]{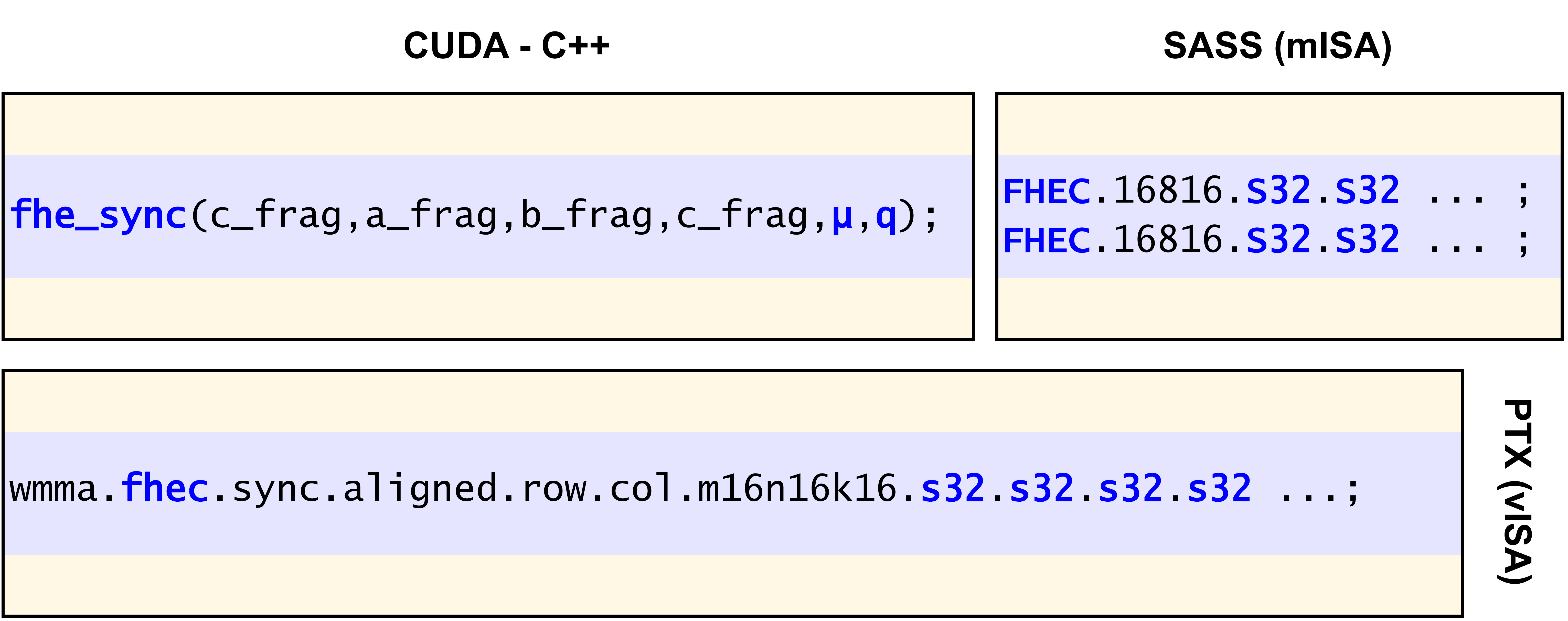}
    \vspace{-20pt}
    \caption{ISA extension for \fhecore: The \texttt{fhe\_sync} intrinsic compiles down to our custom \texttt{FHEC} instruction, executing modulo matrix multiplication on \fhecore.
    The beige sections indicate the reuse of intrinsics and instructions from Figure~\ref{fig:tensor-core-programmability}. 
    The highlighted text indicate changes from TC programming.}
    \label{fig:fhec}
    \vspace{-12pt}
\end{figure}

%% file: tables/tensorcoreextension.tex
\begin{table}
\centering
\caption{RTL metrics when \underline{enhancing tensor cores} to support FHE.}
\vspace{-5pt}
\label{tab:tensorcoreextension}
\resizebox{\columnwidth}{!}{%
\begin{tabular}{clcccc}
\toprule
& & \multicolumn{2}{c}{Enhanced Tensor Core} & \multicolumn{2}{c}{Tensor Core} \\
\cmidrule(lr){3-4} \cmidrule(lr){5-6}
\multicolumn{2}{c}{\centering Metric} & PE & $16\times8$ grid & PE & $16\times8$ grid \\
\midrule
\multirow{3}{*}{} 
Frequency & GHz           & $2.14$    & $1.81$     & $0.76$ - $1.41$    & $0.76$ - $1.41$     \\[2pt]
Latency                          & cycles        & $6$       & $64$ & --       & $64$       \\[2pt]
Area                             & $\mu \text{m}^{2}$ & $10,286.2$ & $115,791$   & $4,954.8$  & $75,577$  \\[2pt]
\midrule
Cumulative area                 & mm$^{2}$      & \multicolumn{2}{c}{\centering $50.01$} & \multicolumn{2}{c}{\centering $32.65$} \\[2pt]
GPU die area & mm$^{2}$ & \multicolumn{2}{c}{\centering $843.36$} & \multicolumn{2}{c}{\centering $826$} \\
\bottomrule
\end{tabular}%
}
\vspace{-10pt}
\end{table}

%% file: src/5-ckks-on-fhecore.tex
\section{CKKS on \fhecore}
\label{sec:ckks-on-fhecore}

In this section, we describe how the core arithmetic kernels of FHE are mapped to the functional units of a GPU equipped with \fhecore{s}.

\subsection{Number Theoretic Transform}
\label{subsec:mapping-number-theoretic-transform}
\input{tables/algorithm}

Just as in TensorFHE~\cite{tensorfhe}, WarpDrive~\cite{warpdrive}, and NEO~\cite{neo}, our primary kernel of interest is the Number Theoretic Transform (NTT).
In \S\ref{subsec:fhe-operations-and-basics-of-ckks} we describe how a large NTT can be formulated into a collection of smaller NTTs, increasing the degree of parallelism, resulting in a better match for GPU acceleration.  
When re-purposing Tensor Cores for computations, the $4$-step NTT can be succinctly expressed as a series of matrix operations, where the shapes of $\text{W}_1$, $\text{W}_2$, and $\text{W}_3$ are $N_2 \times N_1$, $N_1 \times N_1$, and $N_1 \times N_2$, respectively. 
The elements of these matrices are twiddle factors of the form $\psi^{2ij+j}_{(2N_{2}, q)}$, $\psi^{2ij}_{(2N_{1}N_{2}, q)}$ and $\psi^{2ij+i}_{(2N_{1}, q)}$. 
\begin{equation}
\label{eq:attention-modq}
A = \left( (a_{N_1 \times N_2} \times W_1)^T \circ W_2 \right) \times W_3 \bmod q
\end{equation}

The primary challenge in leveraging Tensor Cores for FHE computations lies in their native support for low-precision arithmetic.
Each INT$32/64$ matrix entry in the FHE workload must first be decomposed into consecutive INT$8$ chunks for computation and later recombined for modulo reduction.
These intermediate steps account for up to $40\%$ of the total NTT latency~\cite{warpdrive}.
Moreover, this decomposition dramatically inflates the number of Tensor Core operations.
For matrices with INT$32$ elements, $16$ separate Tensor Core multiplications are required; for INT$64$ values, this number increases to $64$.
Thus, a single \textit{FHECoreMMM} invocation corresponds to $16$ or $64$ \textit{TensorCoreGEMM} calls for functional equivalence.
Algorithm~\ref{alg:ntt} summarizes this procedure for a ``fused'' kernel, where the \textit{NTT\_on\_TensorCore} function performs two rounds of INT$8$ matrix multiplications, interleaved with intermediate recombination and reduction stages that require cross-core communication between Tensor Cores and CUDA cores.

TensorFHE~\cite{tensorfhe} implements this decomposition-based approach by tiling the NTT transform matrix into Tensor Core-compatible dimensions.
The same algorithm, when mapped directly onto \fhecore, can execute without any decomposition or intermediate reduction steps, thereby eliminating the associated (segmentation and recombination) overheads entirely.
To perform a $2^{16}$-point NTT in this manner requires a total of $8192$ \textit{FHECoreMMM} kernel calls.

WarpDrive~\cite{warpdrive}, on the other hand, performs the $4$-step NTT using two levels of decomposition.
This approach reduces the effective matrix size to $16\times16$, allowing the transformation matrices to be directly mapped onto Tensor Cores without any tiling overhead.
With this structure, a $2^{16}$-point NTT requires only $1024$ \textit{FHECoreMMM} kernel calls.

\subsection{Base Conversion}
\label{subsec:mapping-base-conversion}

Base conversion, which is described in \S~\ref{subsec:fhe-operations-and-basics-of-ckks}, transforms the coefficients of a polynomial from one modulus basis to another.
The summation in Equation~(\ref{eq:baseconv}) can be interpreted as a vector dot product, and therefore, computing all residues in the target basis can be represented as a matrix-vector multiplication.
Extending this to all coefficients of the polynomial, the complete process becomes a matrix-matrix multiplication, as shown in Equation~(\ref{eq:base-conversion}). 
\begin{equation}
\resizebox{0.88\columnwidth}{!}{$
\begin{bmatrix} 
\left[ \hat{P}_1 \right]_{q_1} & \hdots & \left[ \hat{P}_{\alpha} \right]_{q_{1}} \\
\vdots & \ddots & \vdots \\
\left[ \hat{P}_1 \right]_{q_{L}} & \hdots & \left[ \hat{P}_{\alpha} \right]_{q_{L}} 
\end{bmatrix}
\begin{bmatrix}
\left[ \hat{P}_1^{-1} a\left[1\right]\left[ 1 \right] \right]_{p_1} & \hdots & \left[ \hat{P}_1^{-1} a\left[ N \right]\left[1\right] \right]_{p_{1}} \\
\vdots & \ddots & \vdots\\
\left[ \hat{P}_{\alpha}^{-1} a\left[ 1 \right]\left[\alpha\right] \right]_{p_{\alpha}} & \hdots & \left[ \hat{P}_{\alpha}^{-1} a\left[ N \right]\left[\alpha\right] \right]_{p_{\alpha}} \\
\end{bmatrix}
\label{eq:base-conversion}
$}
\end{equation}
Unlike NTT, the base-conversion step lacks a recursive structure that can be exploited to reduce its computational complexity.
A second key distinction is that this operation is a ``mixed-moduli'' matrix multiplication because each row of the resulting matrix is reduced under a different modulus.
In current FHE libraries~\cite{FIDESlib,HEonGPU}, both operand matrices are precomputed, and the element-wise multiplication and accumulation are performed on CUDA cores.
On \fhecore, the ``mixed-moduli’’ transformation is handled by programming each column of the systolic array with Barrett constants corresponding to distinct moduli.
This enables partial products to be reduced under different moduli without any modifications to the architecture.
Consequently, the same output-stationary dataflow used for regular modulo matrix transformation can also be applied to base conversion.

The Basis-Aligned Transformation (BAT) optimization in CROSS~\cite{cross} implements base conversion on Google’s TPUv$4$.
However, TPUs (also) lack native support for modulo arithmetic, requiring element-wise postprocessing to compute residue coefficients.
Moreover, their $128\times128$ systolic arrays are significantly larger than what the Base Conversion kernel in FHE workloads demands. 
The number of limbs in a ciphertext polynomial rarely exceeds $30$.
In contrast, \fhecore{’s} $16\times8$ array can tile these matrices more efficiently, avoiding the underutilization penalties incurred while using larger systolic arrays.
Given the right-hand operand matrix contains $N~(\approx2^{16})$ columns, the operation is executed as a sequence of tiled matrix multiplications, each of size $16\times16$.

\input{figures/microarch_side_by_side}

\subsection{Kernels NOT Supported on FHECore}
\label{subsec:kernels-not-supported-on-fhecore}
Apart from NTT and Base Conversion kernels, FHE workloads also include element-wise modulo arithmetic and automorphism operations.

Element-wise modulo additions and multiplications are performed independently on polynomial coefficients.  
These operations are slotwise--they exhibit no inter-element dependencies--and therefore are mapped to the CUDA cores of the GPU, which operate on individual slots in SIMD fashion.

Automorphism applies a permutation to the polynomial coefficients according to a Frobenius map, typically expressed as  
$\pi_r(x) = ([5^r(2x + 1)]_{2N} - 1)/2$ \cite{bts_for_ckks}.  
The automorphism executes in two phases: the first phase performs address generation, computing the destination indices under the Frobenius map, which is mapped onto the CUDA cores; while the second phase performs data rearrangement, handled by the LD/ST units.

\fhecore does not encompass all FHE kernels to execute on it. 
This deliberate separation provides a benefit. 
By delegating element-wise and automorph operations to CUDA cores, while reserving \fhecore for modulo-linear transformations, the GPU can execute both classes of instructions concurrently. 
This concurrency is orchestrated by the warp scheduler, resulting in an increased IPC within each SM for larger workloads.

%% file: tables/algorithm.tex
\begin{algorithm}[!b]
\small
\caption{Tiled NTT on Tensor Cores vs.\ \fhecore}
\label{alg:ntt}
\resizebox{\linewidth}{!}{
\begin{tabular}{|ll|}
\hline
\rowcolor{yellow!25}
\textit{SplitKernel}    & Split INT$32$ entries into four INT$8$ chunks \\
\rowcolor{yellow!25}
\textit{MidKernel}      & Reassemble, reduce, and re-split \\
\rowcolor{yellow!25}
\textit{MergeKernel}    & Final reassembly and modulo reduction \\
\rowcolor{green!25}
\textit{TensorCoreGEMM} & INT$8$ matrix multiplication \\
\rowcolor{lohitpurple!90}
\textit{FHECoreMMM}     & INT$32$ modulo matrix multiplication \\
\hline
\end{tabular}
}
\vspace{0.1em}
\centering
\begin{algorithmic}[1]
  \State \textbf{Function} \textit{NTT\_on\_TensorCore}:
  \For{every $16\!\times\!16$ operand tile pair}
      \State \textit{SplitKernel}      \CUDAComment
      \For{$i = 1$ to $16$}
          \State \textit{TensorCoreGEMM} \TensorComment
      \EndFor
      \State \textit{MidKernel}        \CUDAComment
      \For{$i = 1$ to $16$}
          \State \textit{TensorCoreGEMM} \TensorComment
      \EndFor
      \State \textit{MergeKernel}      \CUDAComment
  \EndFor

  \State \textbf{Function} \textit{NTT\_on\_\fhecore}:
  \For{every $16\!\times\!16$ operand tile pair}
      \State \textit{FHECoreMMM}      \FHEComment
  \EndFor
\end{algorithmic}
\end{algorithm}

%% file: figures/microarch_side_by_side.tex
\begin{figure}
    \centering
    \includegraphics[width=\linewidth]{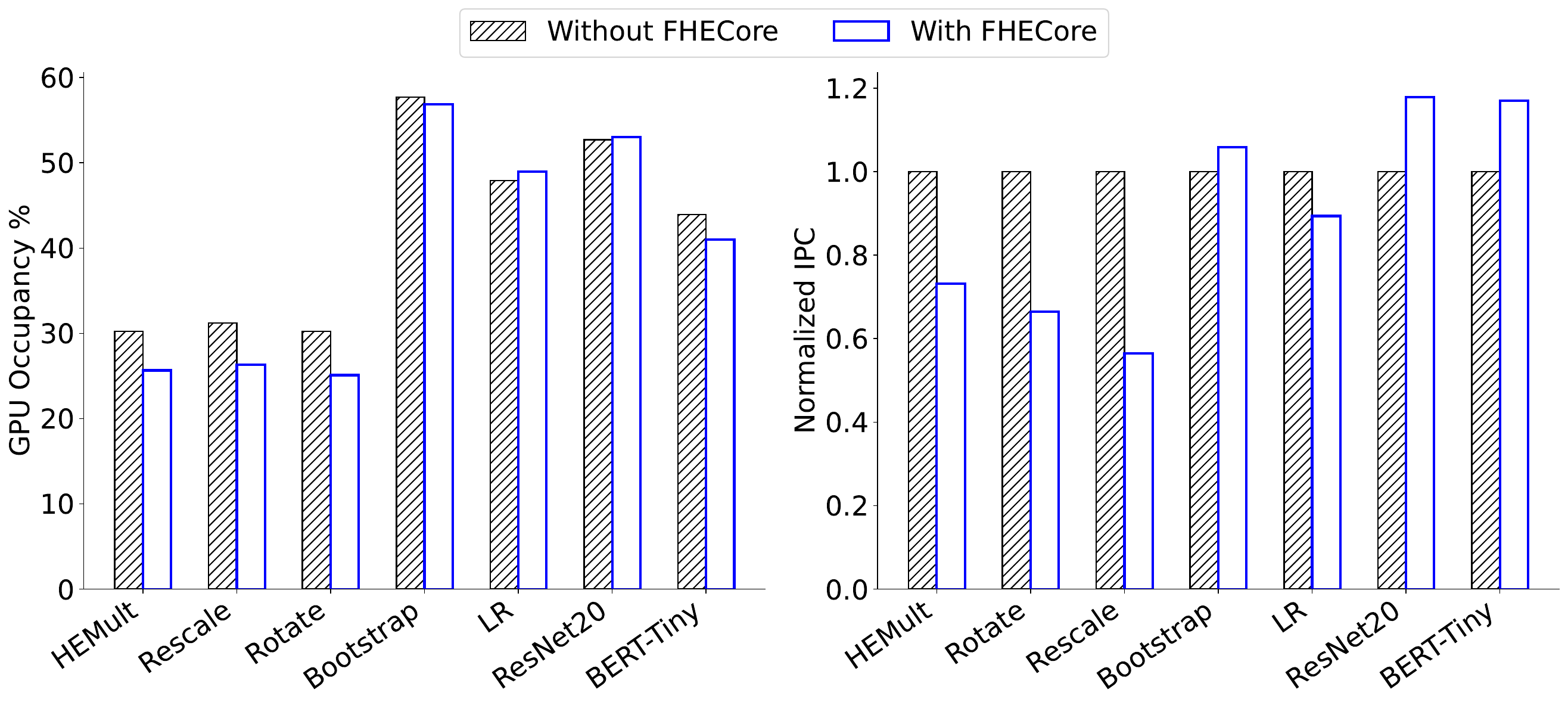}
    \vspace{-20pt}
    \caption{Effect of FHECore integration on GPU occupancy and normalized IPC across CKKS primitives and full workloads.}
    \vspace{-15pt}
    \label{fig:occupancy_ipc}
\end{figure}

%% file: src/6-evaluation.tex
\section{Evaluation}
\label{sec:evaluation}

\input{figures/bootstrap}
\input{figures/result}

We evaluate the impact of integrating \fhecore into the A$100$ GPU -- specifically its effects on performance across CKKS primitives and end-to-end workloads. 
We also report on the resulting reduction in dynamic instruction count and the area overhead of adding \fhecore{s} as independent cores to the A$100$.

\subsection{Methodology}
\label{subsec:methodology}

\input{tables/parameters}

We adopt a simulation-driven evaluation methodology in which FIDESlib applications~\cite{FIDESlib} are executed on a real A$100$ system, and traced using NVBit~\cite{NVBit}. 
These are subsequently replayed on \mbox{Accel-Sim}~\cite{accelsim} in trace-driven mode.
%
We individually profile CKKS primitives -- HEMult, Rescale, and Rotate -- from FIDESlib that execute part of their operations on \fhecore.
For larger workloads, we evaluate full applications, including Bootstrapping, Logistic Regression, ResNet$20$, and BERT-Tiny inference. 
Each program is executed on an NVIDIA A$100$ ($80$~GB) GPU and instrumented using NVBit, which records the SASS instruction issued in each SM, along with their opcodes and GPC/SM identifiers. 
These traces are then simulated on Accel-Sim to assess the microarchitectural impacts of integrating \fhecore.

To incorporate \fhecore into Accel-Sim’s functional model, we introduce our new instruction, \texttt{FHEC}, alongside existing Tensor Core instructions (\texttt{HMMA}, \texttt{DMMA}, \texttt{BMMA}, \texttt{IMMA}). 
This instruction is mapped to `\texttt{SPECIALIZED\_UNIT\_3\_OP}' in Accel-Sim and is assigned a reduced execution latency of $44$ cycles (down from $64$ cycles), which reflects the latency of the output stationary dataflow (discussed in \S\ref{subsec:dataflow}). 
Figure~\ref{fig:occupancy_ipc} shows the impact on the GPU's occupancy and IPC for CKKS primitives and full workloads when this extension is introduced.
Register allocation and dataflow orchestration are handled through the existing WMMA-based Tensor Core mechanism (discussed in \S\ref{subsec:programmability-and-isa-extension}).
Given that the CUDA compilation process is not publicly disclosed, we overcome this challenge by programming \fhecore as if it were a Tensor Core and then replacing all Tensor Core instructions with the \texttt{FHEC} instruction, isolating the benefits of \fhecore without requiring backend compiler modifications.

For area analysis, we implement the \fhecore design in Verilog and synthesize it using the ASAP7~\cite {asap7} ($7$nm) standard-cell library with SiliconCompiler~\cite{siliconcompiler}, which encompasses the complete ASIC flow from RTL to GDS.  

We evaluate performance across four workloads developed using FIDESlib: $(1)$ Bootstrapping, $(2)$ Logistic Regression (LR) training, $(3)$ ResNet$20$ inference, and $(4)$ BERT-Tiny inference.
The CKKS parameters of these workloads can be found in Table~\ref{tab:parameters}.
The LR workload performs training on a downsampled MNIST~\cite{mnist} dataset containing $196$ features.  
The implementation of ResNet$20$ inference is adapted from Rovida et al.~\cite{R20}, with convolution filters encoded as plaintext polynomials.  
Finally, we extend our evaluation to BERT-Tiny~\cite{bert-tiny}, whose parameters are encoded as plaintext polynomials.
It comprises two encoder layers with a dimension of $d = 128$ and $2$ attention heads. 
Matrix multiplications are implemented based on the JKLS technique~\cite{JKLS}.  
Nonlinearities, such as Softmax, LayerNorm, GELU, and $\tanh$, are approximated using Chebyshev expansions and Newton-Raphson iterations.

\subsection{Effect on Bootstrapping}

The impact of \fhecore{} on bootstrapping performance is evaluated through a sensitivity study that sweeps the Fast Fourier Transform (FFT) iteration count from $2$ to $6$.
For each configuration, ideal baby-step and giant-step values were chosen through OpenFHE.
As shown in Figure~\ref{fig:bootstrap}, \fhecore{} reduces both instruction count and latency across all settings, with the largest improvement observed at $\text{FFTIter}=5$.
Under this configuration, the effective bootstrapping time -- defined as total latency divided by the number of limbs remaining after bootstrapping -- drops from $52.3$ ms on the baseline A$100$ to $27.3$ ms with \fhecore.
All values reported in Tables~\ref{tab:programsize} and~\ref{tab:comparison2} correspond to the $\text{FFTIter}=5$ configuration.

\subsection{Improvement in Performance and Reduction in Instruction Count}
\label{subsec:improvement-in-performance}
\input{tables/programsize}
\input{tables/comparison}

Tables~\ref{tab:comparison} and~\ref{tab:comparison2} summarize the end-to-end latency improvements enabled by \fhecore, across representative workloads.
We observe an average (geometric mean) speedup of $1.57\times$ for CKKS primitives mapped to \fhecore and $2.12\times$ for full workloads, with LR and ResNet$20$ showing the largest gains.
To interpret the simulator's cycle count as latency, we assume an average operating frequency of $1087.5$~MHz for the A$100$, based on its dynamic frequency ranging from $765$ to $1410$~MHz.

Similar to prior work~\cite{GME,tensorfhe,warpdrive}, we observe that larger workloads obtain greater speedups than isolated primitives.
We attribute this compounded performance gain to the warp scheduler, which enables both CUDA and \fhecore{s} to execute simultaneously, allowing \texttt{FHEC} instructions and regular instructions to run concurrently. 
This effect explains why larger workloads achieve higher performance gains (Table~\ref{tab:comparison2}) and increased IPC (Figure~\ref{fig:occupancy_ipc}).


Table~\ref{tab:programsize} reports the reduction in dynamic instruction count enabled by the proposed ISA extension.
Across CKKS primitives, we observe a geometric mean reduction of $2.41\times$ in instruction count and a $1.96\times$ reduction for end-to-end workloads.
Figure~\ref{fig:programsizeresult} presents the normalized kernel-level breakdown, illustrating both the relative contribution of each kernel and the compression achieved through the \texttt{FHEC} instruction.

\subsection{Post Synthesis RTL Analysis}
\label{subsec:area-overhead-analysis}

The area and frequency metrics obtained after synthesis with the ASAP$7$ node are summarized in Table~\ref{tab:fhecore_table}.
The frequencies reported denote the maximum achievable clock rates for each individual component. 
All components of \fhecore can operate at a frequency higher than the boosted operating frequency of the A$100$ GPU. 
Hence, they do not fall on the critical path and can be integrated into existing SM pipelines without altering the operating frequency. 
Each Processing Element (PE) is configured with a $6$-cycle latency, and retiming optimizations are enabled in the synthesizer to balance the combinational stages of the pipeline.
Table~\ref{tab:comparisonWithGME} shows that integrating \fhecore into the A$100$ GPU yields a total die size of $845.91~\text{mm}^{2}$--only a $2.4\%$ increase over the $826~\text{mm}^{2}$ baseline and still below the $858~\text{mm}^{2}$ reticle limit~\cite{bigchip}.
By contrast, the GME-integrated MI$100$ grows to $886.2~\text{mm}^{2}$, a $26.6\%$ overhead that exceeds the reticle boundary.
Overall, \fhecore achieves markedly lower area cost than prior GPU-based FHE accelerators.

\input{tables/fhecorertl}


%% file: figures/bootstrap.tex
\begin{figure}
    \centering
    \includegraphics[width=\linewidth]{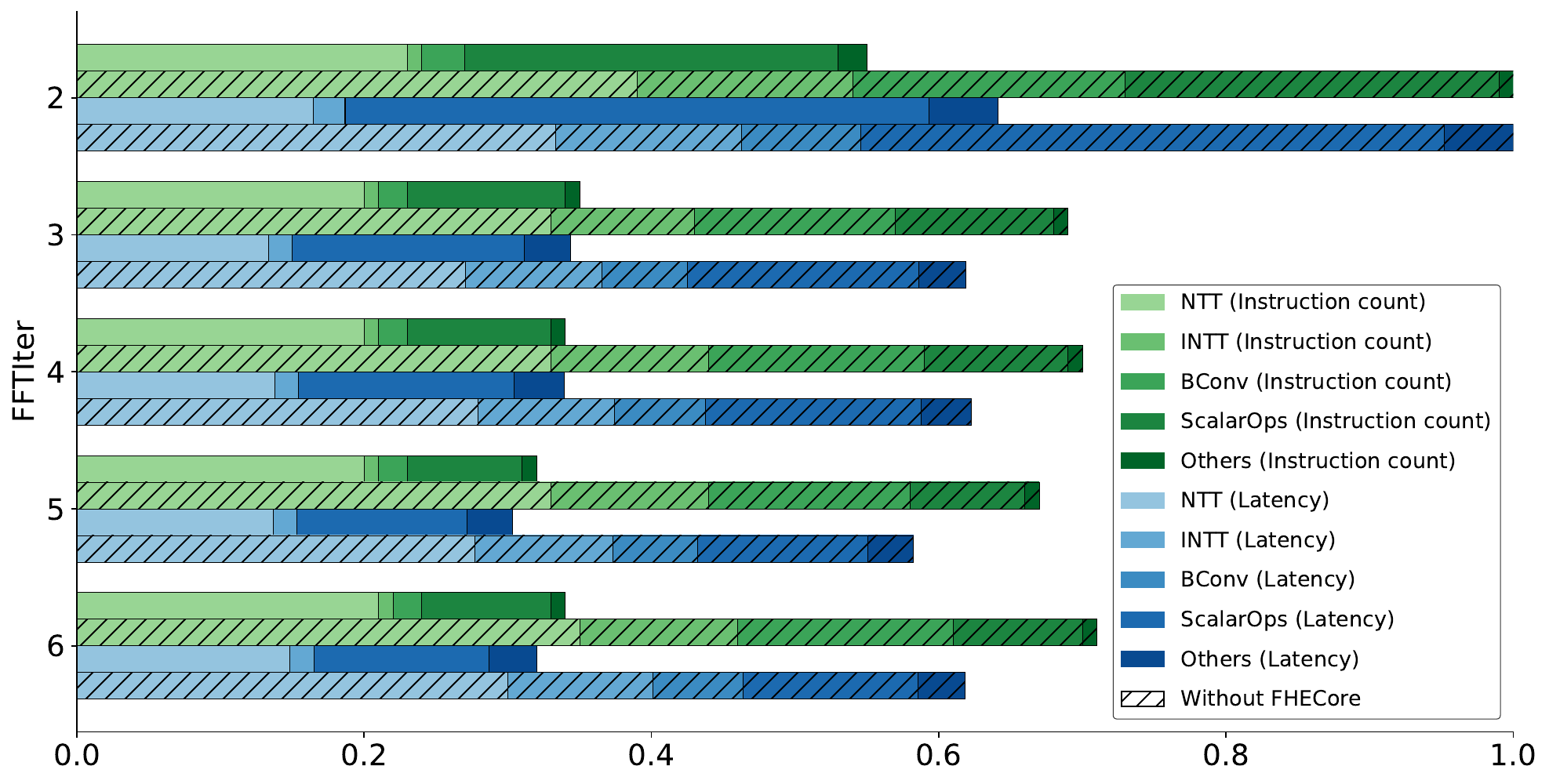}
    \vspace{-20pt}
    \caption{Dynamic instruction count (green) and latency (blue) breakdowns of the bootstrapping kernel across FFT iteration counts, normalized to the FFTIter=2 configuration without \fhecore.
Normalization highlights how both instruction count and runtime vary with and without the proposed \fhecore{} extension.
Hatched bars denote baseline execution without \fhecore; solid bars represent execution with \fhecore.}
    \vspace{-10pt}
    \label{fig:bootstrap}
\end{figure}

%% file: figures/result.tex
\begin{figure*}[!t]
  \centering
  \begin{minipage}{0.48\linewidth}
    \centering
    \includegraphics[width=\linewidth]{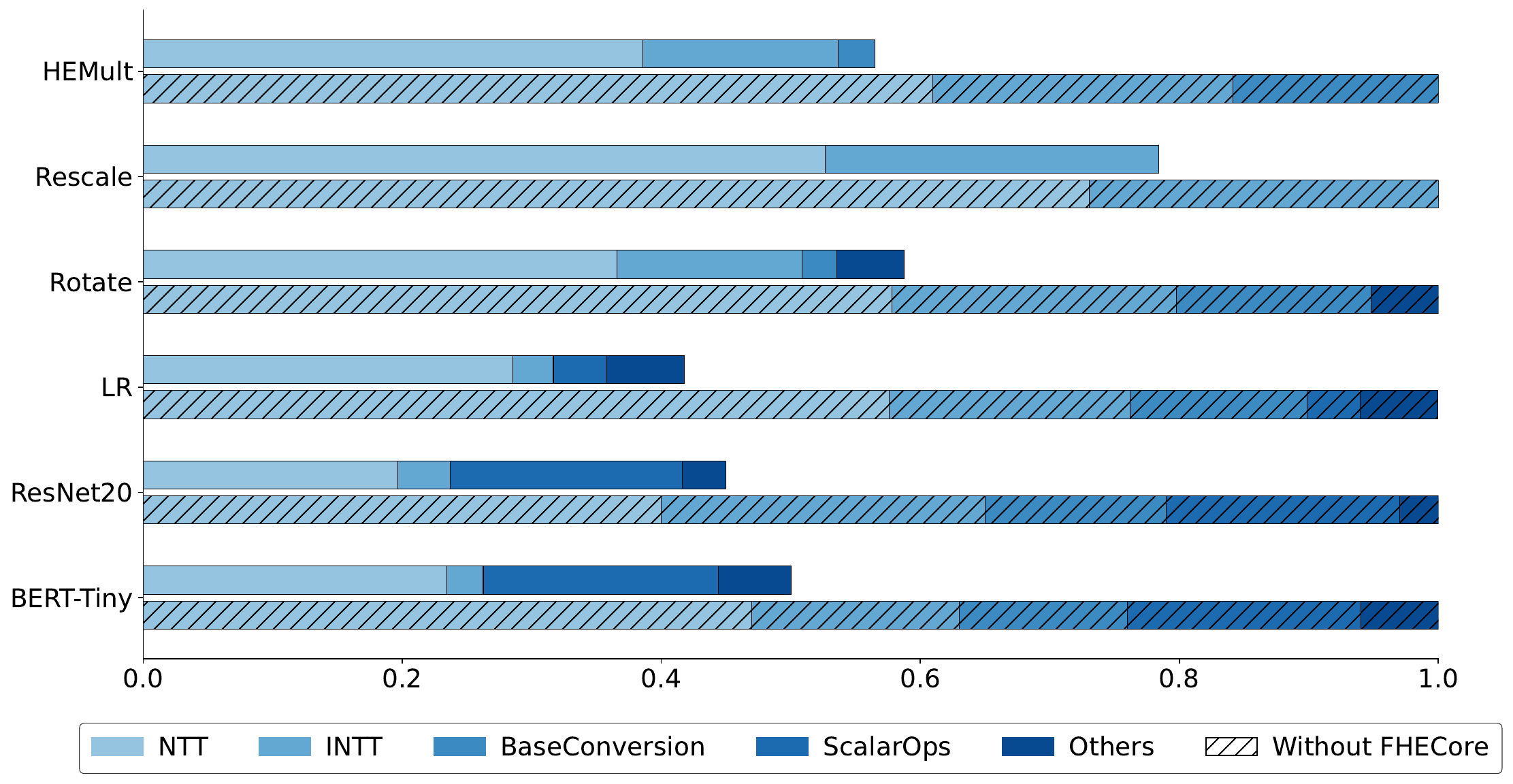}
    \vspace{-20pt}
    \caption{Latency breakdown of different workloads with and without \fhecore in A100 GPU. Absolute values can be found in Table~\ref{tab:comparison} and~\ref{tab:comparison2}.}
    \label{fig:latencyperformance}
  \end{minipage}\hfill
  \begin{minipage}{0.48\linewidth}
    \centering
    \includegraphics[width=\linewidth]{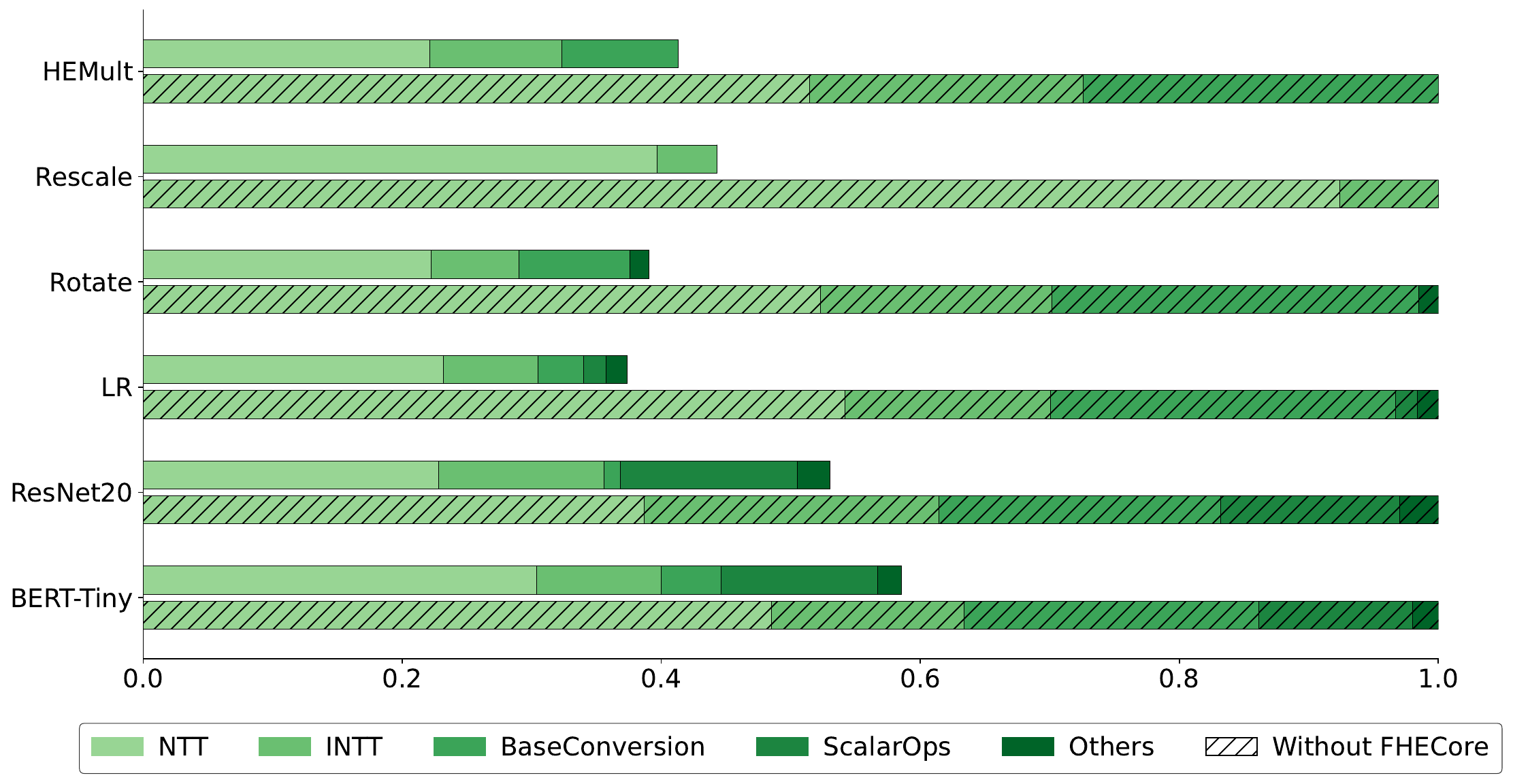}
    \vspace{-20pt}
    \caption{Breakdown of dynamic instruction count for different workloads with and without \fhecore in A100 GPU. Absolute values can be found in Table~\ref{tab:programsize}.}
    \label{fig:programsizeresult}
  \end{minipage}\hfill
  
  \vspace{-10pt}
\end{figure*}


%% file: tables/parameters.tex
\begin{table}
\centering
\caption{CKKS parameters for the benchmarks.}
\vspace{-5pt}
\label{tab:parameters}
\resizebox{0.85\linewidth}{!}{%
\begin{tabular}{lcccccc}
\toprule
Benchmark& $\lambda $& $log N $ & $log QP $ & $L $ & $L_{eff} $ & $dnum $ \\
\midrule
Bootstrap & $128$ & $16$ & $1743$ & $26$ & $6$ & $3$ \\[2pt]
LR &$128$ & $16$ & $1675$ & $29$ & $6$ & $4$ \\[2pt]
ResNet$20$ &$128$ & $16$ & $1714$ & $26$ & $8$ & $4$ \\[2pt]
BERT-Tiny &$128$ & $16$ & $1740$ & $26$ & $7$ & $5$ \\
\bottomrule
\end{tabular}%
}
\vspace{-10pt}
\end{table}

%% file: tables/programsize.tex
\begin{table}
\centering
\caption{Reduction in Dynamic Instruction Count with \fhecore.}
\vspace{-5pt}
\resizebox{0.9\columnwidth}{!}{
\begin{tabular}{lccl}
\toprule
 & A100 & A100 with \texttt{FHEC} \\
\midrule
HEMult   & $139,449,088$ & $57,604,096$&$(\mathbf{2.42\times})$ \\[4pt]
Rotate   & $146,941,952$ & $57,383,936$&$(\mathbf{2.56\times})$ \\[4pt]
Rescale  & $29,974,528$ & $13,278,720$&$(\mathbf{2.26\times})$ \\[4pt]
Bootstrap & $36,129,286,144$ & $17,073,057,143$&$(\mathbf{2.12\times})$ \\[4pt]
LR & $89,385,078,641$ & $33,384,095,353$&$(\mathbf{2.68\times})$ \\[4pt]
ResNet & $556,714,062,191$ & $295,196,429,566$&$(\mathbf{1.89}\times)$ \\[4pt]
BERT-Tiny & $1,809,067,028,156$ & $1,059,032,787,283$&$(\mathbf{1.71}\times)$\\
\bottomrule
\end{tabular}
}
\label{tab:programsize}
\vspace{-10pt}
\end{table}

%% file: tables/comparison.tex
\begin{table}
\centering
\caption{Latency ($\mu s$) comparison with other GPU works. \\[0pt]
Implementation publicly available? ~~~ \green{\cmark} Yes ~~~ \red{\xmark} No}
\vspace{-5pt}
\label{tab:comparison}
\resizebox{0.92\columnwidth}{!}{
\begin{tabular}{lcccc}
\toprule
\textbf{~} & & Rescale & Rotate & HEMult \\
\midrule
{\green{\cmark} OpenFHE~\cite{OpenFHE}}        & CPU ($24$ threads) & $4920$  & $105300$ & $151580$ \\[4pt]
{\green{\cmark} Phantom~\cite{PhantomFHE}}     & RTX$4090$          & $224$   & $1139$  & $1220$  \\[4pt]
{\red{\xmark} TensorFHE~\cite{tensorfhe}}      & RTX$4090$          & $115$   & $18592$ & $18689$ \\[4pt]
{\red{\xmark} Neo~\cite{neo}}                  & A$100$             & $114$   & $3422$  & $3472$  \\[4pt]
{\red{\xmark} Cheddar~\cite{Cheddar}}     & RTX$4090$          & $68$      & $476$      & $533$      \\[4pt]
{\green{\cmark} HEonGPU~\cite{HEonGPU}}        & RTX$4090$          & $150$   & $8200$  & $8172$  \\[4pt]
{\green{\cmark} FIDESlib~\cite{FIDESlib}}      & RTX$4090$          & $156$   & $1107$  & $1084$  \\
\midrule
{~~ FIDESlib}                                  & A$100$ (Baseline)  & $227$   & $1261$  & $1196$  \\[4pt]
{~~ FIDESlib}                                  & A$100$ + \fhecore & $178$ & $741$ & $675$  \\
                                               &                    & (\textbf{1.28$\times$}) & (\textbf{1.70$\times$}) & (\textbf{1.77$\times$}) \\
\bottomrule
\end{tabular}
}
\vspace{-10pt}
\end{table}
\begin{table}
\centering
\caption{Latency ($ms$) comparison of end-to-end workloads \\between A100 and A100 + \fhecore systems.}
\vspace{-5pt}
\label{tab:comparison2}
\resizebox{0.9\columnwidth}{!}{%
\begin{tabular}{@{}lcccc@{}}
\toprule
 & Bootstrap & LR & ResNet$20$ & BERT-Tiny \\
\midrule
A$100$ (Baseline) & $314.67$  & $747.44$  & $5028.23$ & $16583.83$ \\[4pt]
A$100$ + \fhecore & $163.90$ & $312.37$ & $2262.16$ & $8300.38$ \\
 & $(\mathbf{1.92\times})$ & $(\mathbf{2.39}\times)$ & $(\mathbf{2.22}\times)$ & $(\mathbf{2}\times)$ \\
\bottomrule
\end{tabular}
}
\vspace{-10pt}
\end{table}

%% file: tables/fhecorertl.tex
\begin{table}
\centering
\caption{RTL Metrics of \fhecore}
\vspace{-5pt}
\label{tab:fhecore_table}
\begin{tabular}{lccc}
\toprule
 & $~~~~~~~~~~~~~~~~~~~~~~~~$ & \multicolumn{2}{c}{\fhecore} \\
\cmidrule(lr){3-4}
\multicolumn{2}{c}{Metric} & PE & $16\times8$ grid \\
\midrule
\textsuperscript{\textdaggerdbl}Frequency & GHz    & 3.50     & 1.58     \\[4pt]
Latency             & cycles & 6        & 44       \\[4pt]
Area                & $\mu \text{m}^{2}$ & 5,901.1   & 46,096.5  \\
\midrule
Cumulative area     & mm$^2$ & \multicolumn{2}{c}{19.91} \\
\bottomrule 
\end{tabular}
\vspace{-10pt}
\end{table}
\begin{table}
\centering
\caption{Comparison of area overhead with prior work.}
\label{tab:comparisonWithGME}
\vspace{-5pt}
\begin{tabular}{@{}lcc@{}}
\toprule
 & Shivdikar et al.~\cite{GME} & This Work \\ 
\midrule
\multirow{2}{*}{Architectural baseline}
  & AMD MI$100$        & NVIDIA A$100$      \\[2pt]
  & $700~\text{mm}^{2}$& $826~\text{mm}^{2}$\\[4pt]
\multirow{1}{*}{Modified GPU area}
  & $886.2~\text{mm}^{2}$ & $845.91~\text{mm}^{2}$ \\[0pt]
  \midrule
\multirow{1}{*}{Overhead incurred}
  & $\red{+26.6\%}$    & $\green{+2.4\%}$   \\
\bottomrule
\end{tabular}
\vspace{-12pt}
\end{table}

%% file: src/7-discussion.tex
\section{Discussion}
\label{sec:Discussion}


\textit{Are there other (non-CKKS) workloads that could benefit from this functional unit?} FHE schemes like BFV and BGV share CKKS’s underlying polynomial arithmetic; their main differences stem from the encoding layer, which tailors them to different application domains. Consequently, they rely on the same dominant kernels--Number Theoretic Transform and base conversion--both of which map naturally onto \fhecore.
TFHE traditionally employs FFTs because it's polynomial coefficients may be integers or floating-point values, but with appropriate modulus selection, these FFTs can be realized as NTTs--a strategy already used in OpenFHE~\cite{OpenFHE} and NuFHE~\cite{nufhe}--making TFHE’s polynomial components equally compatible with \fhecore.

Beyond FHE, the functional unit is broadly applicable to a wide range of cryptographic workloads that rely on modulo arithmetic. 
This includes Zero-Knowledge Proof (ZKP) systems, which, like lattice-based cryptography, involve heavy use of modular matrix-matrix multiplications and NTT-based operations~\cite{zkprophet}. 
In such settings, \fhecore{} can act as a high-throughput modulo arithmetic engine, effectively paving a path toward a unified high performance accelerator for cryptography~\cite{UCA} capable of handling diverse privacy-preserving protocols efficiently on GPUs.

\vspace{7pt}
\textit{Are there workloads that take a hit in performance because of this functional unit?}
Since \fhecore~introduces microarchitectural modifications to the GPU, it is reasonable to question if these changes introduce inefficiencies for workloads that do not use the functional unit.
We would like to emphasize that all architectural decisions discussed in \S~\ref{sec:fhecore} were made carefully to ensure that the existing SM pipeline remains unaffected. 
The only potential performance conflict arises when a workload requires concurrent use of both the tensor core and \fhecore. 
However, since plaintext machine learning and FHE are fundamentally disjoint workloads, we are unaware of any real-world applications that exhibit this usage pattern. 
In contrast, GME introduces routers at the L$2$ cache, adding pipeline stages that impact all workloads, specifically ones that aren't able to take advantage of the network-on-chip. 
\fhecore~avoids such system-wide modifications, ensuring no disruption to existing GPU workloads.

\vspace{7pt}
\textit{Applicability to GPUs beyond A$100$}: While our evaluations and integration strategy are centered on the NVIDIA A$100$ GPU, the design of \fhecore is not tied to this specific architecture. 
The functional unit and its programming model can be readily extended to newer architectures such as Hopper H$100$ and Blackwell B$100$. 
For GPUs lacking dedicated Tensor Cores, \fhecore~can be instantiated as a standalone unit. 
A coarse estimate of the area overhead of \fhecore~on H$100$/B$100$ is $1.5\%$, which remains comfortably within the reticle limit.
These newer architectures also introduce Tensor Memory, a dedicated on-chip store for matrix tiles accessed via Tensor Memory Accelerators (TMAs)~\cite{nvidia_hopper}, rather than through the register file as in Ampere.
Adapting \fhecore to work with Tensor Memory requires only one modification: the functional unit would interface with Tensor Memory instead of the register file, leaving the rest of the architecture and programming model intact.

\vspace{7pt}
\textit{Can other GPU libraries for FHE take advantage of \fhecore?}
In \S~\ref{subsec:programmability-and-isa-extension}, we expose \fhecore's programmability up to the MMA intrinsics level via \texttt{fhe\_sync}, analogous to how tensor core intrinsics are used by various ML libraries. 
These intrinsics can be invoked by GPU-based FHE libraries, enabling seamless integration with existing kernels. 
Since \fhecore~operates at the granularity of $16 \times 16$ modulo matrix multiplications, it can readily adapt to evolving algorithmic changes in FHE schemes.


%% file: src/8-related-work.tex
\section{Related Work}
\label{sec:related-work-and-discussion}




Over the years there have been several works that explore opportunities to accelerate FHE.
Starting with efficient CPU implementations which incorporate advancements in the algorithmic front of FHE~\cite{sealcrypto, OpenFHE}. 
However due to the limited potential for parallelism in CPUs their adoption is not practical for real-world encrypted computing workloads.
ASIC accelerators~\cite{bts,craterlake,f1, ARK, cinnamon, sharp} provide the best raw performance, but they are expensive--in terms of cost, time and verification--therefore aren't flexible to the rapid evolution of FHE.
Likewise, FPGA accelerators provide reconfigurable architectures that can adapt to changing schemes and parameters; however, they are limited by their clock speeds and low device memories~\cite{FAB, HEAX, HEAP, book}.

GPU-based accelerators show tremendous success in accelerating FHE operations.  
TensorFHE~\cite{tensorfhe}, Warpdrive~\cite{warpdrive} and Neo~\cite{neo}, specifically focus on utilizing Tensor Cores to accelerate time-consuming Number Theoretic Transform (NTT) operations of CKKS FHE.
Phantom~\cite{PhantomFHE} provides GPU implementations of not only the CKKS scheme, but also BGV and BFV in one unified framework. 
These schemes are implemented with kernel fusing and data reuse optimizations and detailed benchmarking.
Cheddar~\cite{Cheddar} achieves remarkable speedups over other GPU-based FHE frameworks by adopting double-rescaling and $32$-bit wide datatypes only. 
On a related direction, CROSS~\cite{cross} explores the viability to run CKKS programs on Google's TPUs, which are systolic arrays designed for matrix multiplications in ML.
There have also been heterogeneous architectures with system level modifications such as--Anaheim~\cite{anaheim} and MemFHE~\cite{MemFHE}--that use near- or in-memory computing approaches to address the memory boundedness of these workloads.



%% file: src/9-conclusion.tex



\section{Conclusion}

This work shows that contemporary GPUs, despite their abundant parallelism, remain fundamentally mismatched to the wide-precision modulo arithmetic that dominates FHE. By recognizing that Number Theoretic Transform and Base Conversion are key bottlenecks and can be expressed as modulo-linear transformations, we introduce \fhecore, a systolic unit that performs wide-precision modulo multiply–accumulate with integrated Barrett reduction, eliminating the long chains of fine-grained CUDA-core instructions.

Across CKKS workloads, \fhecore reduces the dynamic instruction count by (geometric-mean) $2.41\times$ for primitives and $1.96\times$ for full applications, while improving performance by $1.57\times$ and $2.12\times$, respectively -- including a $50\%$ reduction in bootstrapping latency -- all with only $2.4\%$ area overhead. 
The design integrates cleanly within existing GPU toolchains and generalizes to newer architectures, offering a practical, scalable path to high-performance encrypted computation.




%% file: src/acknowledgements.tex
\section*{Content Generated by AI}

We used ChatGPT and Google Gemini strictly for language refinement, limited to grammar, clarity, and phrasing suggestions. All technical ideas, designs, simulations, and results were fully conceived and validated by us, and no AI generated text was included without our careful verification.

\section*{Acknowledgements}
This work was supported in part by NSF CNS 2312275 and 2312276.